\documentclass[a4paper,10pt,twoside]{cpc-hepnp}
\usepackage{CJK,upgreek,fancyhdr}
\usepackage{multicol}
\usepackage{graphicx}
\usepackage{booktabs}
\usepackage{amssymb,bm,mathrsfs,bbm,amscd}
\usepackage[tbtags]{amsmath}
\usepackage{lastpage}
\pdfoutput=1
\usepackage[T1]{fontenc}
\usepackage{float}
\usepackage{amsbsy}
\usepackage{amsmath}
\usepackage{amstext}
\usepackage{hyperref}
\usepackage{graphicx}
\usepackage{esint}
\usepackage{subfigure}
\usepackage{upgreek}
\usepackage{wrapfig}
\usepackage{amscd}
\usepackage{xcolor}
\usepackage{hyperref}
\usepackage{overpic}
\graphicspath{{pic/}}

\begin{document}

\fancyhead[c]{\small Submitted to Chinese Physics C}
\fancyfoot[C]{\small 010201-\thepage}


\title{Investigating the topological structure  of quenched lattice QCD with overlap fermions by using multi-probing approximation\thanks{Supported in part by the National Natural Science Foundation of China (NSFC) under the project No.11335001, No.11275169, No.11075167. It is also supported in part by the DFG and the NSFC (No.11261130311) through funds provided to the Sino-German CRC 110 "Symmetries and the Emergence of Structure in QCD". This work was also funded in part by National Basic Research Program of China (973 Program) under code number 2015CB856700.} }

\author{%
      You-Hao Zou $^{1;1)}$\email{11006067@zju.edu.cn}%
\quad Jian-Bo Zhang $^{1;2)}$\email{jbzhang08@zju.edu.cn}%
\quad Guang-Yi Xiong $^{1;3)}$\email{xionggy@zju.edu.cn}%
\\
\quad Ying Chen$^{2,3}$
\quad Chuan Liu$^{4,5}$
\quad Yu-Bin Liu$^{6}$
\quad Jian-Ping Ma$^{7}$
}
\maketitle

\address{%
$^1$ Department of Physics, Zhejiang University, Zhejiang 310027,
P.R. China\\
$^2$ Institute of High Energy Physics, Chinese Academy of Sciences, Beijing 100049, China\\
$^3$ Theoretical Center for Science Facilities, Chinese Academy of Sciences, Beijing 100049,P.R. China \\
$^4$ School of Physics, Peking University, Beijing 100871, P.R. China\\
$^5$ Collaborative Innovation Center of Quantum Matter, Beijing 100871, P.R. China\\
$^6$ School of Physics, Nankai University, Tianjin 300071, P.R. China \\
$^7$ Institute of Theoretical Physics, Chinese Academy of Sciences,Beijing 100080, P.R. China
}

\begin{abstract}
The topological charge density and topological susceptibility are determined by multi-probing approximation using overlap fermions in quenched SU(3) gauge theory. Then  we  investigate the topological structure of the quenched QCD vacuum, and compare it with results from the all-scale topological density, the results are consistent. Random permuted topological charge density is used to check whether these structures represent underlying ordered properties. Pseudoscalar glueball mass is extracted from the two-point correlation function of the topological charge density. We study $3$ ensembles of different lattice spacing $a$ with the same lattice volume $16^{3}\times32$, the results are compatible with  the results of all-scale topological charge density, and the topological structures revealed by multi-probing are much closer to all-scale topological charge density than that by eigenmode expansion.
\end{abstract}

\begin{keyword}
topological structure,  topological susceptibility, multi-probing approximation, cluster
\end{keyword}

\begin{pacs}
11.15.Ha, 12.38.-t, 12.38.Gc
\end{pacs}

\footnotetext[0]{\hspace*{-3mm}\raisebox{0.3ex}{$\scriptstyle\copyright$}2013
Chinese Physical Society and the Institute of High Energy Physics
of the Chinese Academy of Sciences and the Institute
of Modern Physics of the Chinese Academy of Sciences and IOP Publishing Ltd}%

\begin{multicols}{2}
\section{Introduction}

The topological susceptibility $\chi$ of Yang-Mills theory is an important physical quantity in explaining
the $U(1)_{A}$ anomaly  and the mass of $\eta^\prime$, e.g. the study of this problem in
 $N_{c}\rightarrow \infty$ limit brings up the famous Witten-Veneziano formula for $\eta^{\prime}$ mass~\cite{WittenVeneziano1, WittenVeneziano2, WittenVeneziano3}.
 In the continuum limit, the topological susceptibility as a function of four-momentum $k$,
 denoted by $\chi(k^2)$ in Euclidean version, is defined as
\begin{equation}\begin{split}\label{chi}
    \chi(k^2) & =  \int d^4x ~~\exp(ikx)C_{qq}(x) \\
    & = \int d^4x ~\exp(ikx)\langle 0|T(q(x)q(0))|0\rangle,
\end{split}\end{equation}
here the quantity $C_{qq}(x) = \langle 0|T(q(x)q(0))|0\rangle $ ~is the two-point correlator of the topological charge density $q(x)$ and $|0\rangle$ denotes the QCD vacuum.

The contribution of the $U(1)_{A}$ anomaly to the $\eta^{\prime}$ mass is given by~\cite{WittenVeneziano1, WittenVeneziano2, cqq1}
\begin{equation}\label{chi2}
    \frac{F^{2}_{\pi} m^2_{\eta^{\prime}}}{2N_{f}} = \chi^{YM}(0),
\end{equation}
in which $YM$ is for pure Yang-Mills theory, $N_f$ is the number of flavors, and $F_{\pi}$ is the corresponding $\pi$ decay constant, and will leads to $\chi^{YM}(0) \approx (180\textrm{MeV})^{4}$~\cite{WittenVeneziano2, cqq1}.

To calculate $q(x)$ and $\chi(0)$  of  SU(3) pure Yang-Mills theory on lattice, we have to define the topological density $q(x)$ first. There are several ways  to define  it.
One usually way is to define $q(x)$  with the field strength tensor $F_{\mu\nu}$ on the lattice as~\cite{qx:fmn}
\begin{equation}\label{qlx}
    q_{L}(x) = -\frac{1}{32\pi^2}\sum_{\mu\nu\rho\sigma} \epsilon_{\mu\nu\rho\sigma} Tr{ F_{\mu\nu}(x) F_{\rho\sigma}(x) },
\end{equation}
in which subscript $L$ denotes lattice.
This definition has  the  classical continuum limit $ q_{L}(x) \xrightarrow{a\rightarrow 0} q(x)a^4 $. But on the lattice, since the direct calculated topological charge $Q_{L} = \sum_{x} q_{L}(x)$ usually deviates from
the integral values, and $Q_{L} = a^{4}Z(\beta)Q + \mathcal{O}(a^{6})$~\cite{kiinLQCD, kiprimeinLQCD1, kiprimeinLQCD2}, with the multipliciative renormalization coefficient $Z(\beta)$ dependent on the inverse coupling constant $\beta$.
This dependence can be computed perturbatively, or some kind of smoothing steps could be used to make the effective inverse coupling constant $\beta_{\rm eff} $ larger and such that
$Z(\beta_{eff}) \rightarrow 1 $~\cite{kiprimeinLQCD2} is achieved. As to $\chi(0)$, it was found that multiplicative and additive renormalization terms both exist  as $\displaystyle\chi_{L}(0) = a^{4} Z^2(\beta)\chi(0) + a^{4} A(\beta)<S(x)>_{n.p.}+P(\beta)$~\cite{kiinLQCD}, $S(x)$ is the action density, n.p. indicates non-perturbative, $A(\beta)$ and $P(\beta)$ are ultra-violet effects that can be approximated in perturbation theory. We can also smooth configurations to approach integer value of $Q_{L}$ for calculating $\chi(0)$. But the smooth procedure would influence $q(x)$.

Another way to define $q(x)$ is using massless Dirac operator $D(0,x,y)$ that obeys the Ginsparg-Wilson relation $D\gamma_5  + \gamma_5 D = aD\gamma_5 D$. Here the first argument in $D(m,x,y)$ denotes the mass of the quark while
 the second and third index are the space-time index on the lattice. For simplicity, we will also use
 $D(0)$ to denote the massless Dirac operator with the space-time indices suppressed. With this notation,
 the topological charge density can be defined as
\begin{equation}\label{qx:overlap}
    q_L(x) = \frac{1}{2} \texttt{tr}_{cd}\gamma_5 D(0,x,x),
\end{equation}
where $c$ and $d$ being the index of color and Dirac spinor respectively.
This is a proper approach that guarantees the topological charge $Q_L$  an integral value
since Ginsparg-Wilson fermion satisfies Atiyah-Singer index theorem~\cite{qx:trQ}
\begin{equation}\label{index}
    Q_{L} =\sum_{x} q_{L}(x) = \frac{1}{2} \texttt{tr}\gamma_5 D(0) = n_L - n_R,
\end{equation}
in which $n_{L}$ and $n_{R}$ are number of left- and right-handed zero modes of $D(0)$.
Massless overlap fermion is exactly Ginsparg-Wilson fermion~\cite{overlap1, overlap2}, so in this paper we will use massless overlap Dirac operator $D_{ov}(0)$ to calculate the topological charge density.
The definition of $D_{ov}(0)$ reads,
\begin{equation}\label{eq:overlap}
    D_{ov}(0)  = \frac{1}{a}(1+\gamma_{5}{\rm sign}(H)) ,~~~~H = \gamma_{5}D_{w}(s),
\end{equation}
$D_{w}(s)$ is the Wilson-Dirac operator, with parameter $|s| < 1$ for the optimization of the locality of overlap operator~\cite{localityofoverlap1}, $D_{w}(s) = D_{w} - (1+s) $, $D_{w}$ is the massless Wilson-Dirac operator
 with mass parameter $m=-(1+s)$. In this work we choose parameter $\kappa=\frac{1}{2(am+4)} = 0.21$, which means  $s=0.619$.

Nevertheless, it will cost too much computer resources to calculate topological charge density using the definition in Eq.~\eqref{qx:overlap} directly, i.e., using point sources to calculate the trace, the resulting topological charge density $q(x)$ is named all-scale topological charge density in Ref.~\cite{qx:structure3}. An approximation method called eigenmode expansion was proposed in the literature~\cite{qx:structure1, qx:structure2, qx:structure3}:
$ q_{\lambda_{cut} }(x) = - \sum_{|\lambda| < \lambda_{cut}} (1-\frac{\lambda}{2})p_{\lambda_{5}}(x)$, $ p_{\lambda_{5}}(x) = \sum_{cd} \psi_{\lambda}^{\dagger}(x) \gamma_{5} \psi_{\lambda}(x)$, $\psi_{\lambda}$ is the eigenmode of $D(0)$ with eigenvalue $\lambda$, $\lambda_{cut}$ is the supremum of the absolute eigenvalue of the corresponding eigenmodes included in the eigenmode expansion.

Only if all zero modes  are included, which obey the relationships $\gamma_{5}\psi_{\lambda} = \pm \psi_{\lambda},~D_{ov}(0)\psi_{\lambda} = 0 $, $q_{\lambda_{cut} }(x)$ would  sum the individual local chiralities, and the topological charge of eigenmode expansion $Q_{\lambda_{cut} } = Q_{L} $. However, some features such as the positive core of two-point correlator of topological charge density $C_{qq}(x)$  would increases, and the height of peaks of $C_{qq}(x)$ would decrease, even the negative region of $C_{qq}(x)$ would disappear when $\lambda_{cut}$ is too small~\cite{qx:structure3}, therefore more eigenmodes should be considered which would also cost too much computer time, and it's not sure what has been discarded in the  eigenmodes of larger $\lambda_{cut}$.

In this paper we will use multi-probing method~\cite{multi-probing} to approximate $q(x)$ owing to the finite range of $D(0)$~\cite{localityofoverlap, qx:structure3}, $\delta Q$, the deviation of $Q_{L}$ from an integer would be regarded as a parameter that signals the validity of this approximation.
The lattice version of Eq.~\eqref{chi} is defined as
\begin{equation}\label{chiLQCD}
    \chi_{L}(k^2) = \sum_{x \in V} \exp(ikx) C_{qq}(x),
\end{equation}

with the lattice version of two-point correlation function $C_{qq}(r)=\displaystyle{\frac{1}{V}\sum_{x \in V}\langle q(x+r)q(x)\rangle}$.
Noting that the true value of expectation of topological charge is 0, the lattice version of topological susceptibility at $k^{2}=0$  can be expressed as
\begin{equation}\label{chi0LQCD}
    \chi_{L}(0) = \frac{\langle Q_{L}^{2}\rangle}{V},
\end{equation}
obviously it has the same accuracy of topological charge.

It is believed that the $U(1)_{A}$ problem~\cite{U1problem}, topological susceptibility, $\eta^{\prime}$ mass, $\theta$ dependence and spontaneous chiral symmetry breaking are all related to the vacuum fluctuation of topological charge density~\cite{Witten-instanton, WittenVeneziano1}. So in Sect.~\ref{fluctuation} we  attempt  to investigate the local topological structures of the quenched QCD vacuum, hoping that this will reflect some ordered features of the quenched QCD vacuum. Since they can be explained by instantons model, which was first introduced by Polyakov~\cite{instantons1,instantons2}, it is reasonable to try to describe the structure of topological charge density in terms of instanton liquid  model, the topological charge density shows sign-coherent unit-quantized lumps locally. In the past, several papers~\cite{qx:instantons1, qx:instantons2, qx:instantons3, qx:instantons4,
qx:instantons5, qx:instantons6, qx:instantons7, qx:instantons8, qx:instantons9} have attempted to justify this idea, but these papers used methods such as cooling, smoothing and smearing, which will lead to configurations towards classical solutions, therefore the results should be taken with some reservations. On the other hand, it was predicted that QCD vacuum doesn't resemble the instanton liquid model~\cite{Witten-instanton}, and this idea was further supported by lattice evidence in Ref.~\cite{qx:structure1}. In fact, there is no four dimension global structure of topological charge density but only lower dimensional structures survive~\cite{qx:structure1, qx:structure2, qx:structure3}. In Sect.~\ref{fluctuation} we will discuss more details about the structure of topological charge density, and the results of all-scale $q(x)$ in Ref.~\cite{qx:structure3} will be compared with, which were built up from three ensembles only with $53$, $5$ and $2$ configurations, due to the large amount of computer time for calculating all-scale topological charge density.

\section{Multi-probing approximation}~\label{app:mp}

The massless overlap operator $D_{ov}(0)$ is not ultralocal but local.
It is exponentially decaying with respect to lattice distance with a decay rate that has little
independent on inverse coupling constant $\beta$~\cite{localityofoverlap, qx:structure3}
\begin{equation}\label{D(0,x,y)}
     D_{ov}(0,x,x+r) \propto \exp(-\mu r/a) ~~,~~\mu > 0,
\end{equation}
in which $r$ is the lattice Euclidean distance, $a$ is the lattice spacing, so we have $D_{ov}(0,x,x+r)\ll D_{ov}(0,x,x)$ when $r/a$ is large enough. Therefore we can approximate topological charge density $q(x)$ with massless overlap fermions as
\begin{equation}\begin{split}\label{qx:mp}
    q_{mp}(x) & = \frac{1}{2}\textrm{tr}_{cd} \gamma_{5} (D_{ov}(0,x,x) \\
              & + D_{ov}(0,x,x+r_1) + D_{ov}(0,x,x+r_2) \\
              & + D_{ov}(0,x,x+r_3) + \cdots),
\end{split}\end{equation}
in which $\{r_{i}\}$ are some distances satisfied $ exp(-\mu r_{i}/a ) \ll 1$, and $c, d$ are indices for Dirac and color respectively.

In actual computing, we choose sites on the lattice $\{x_{i}\}$ as a group of probes called multi-probe
and these sites are distributed uniformly with interval $\delta r $, $\delta r$ should be an integer number that is a common divisor of space-time lengths of the lattice, or multiplied with $\sqrt 2$ when the sites in the same group are separated by odd and even into two new groups~\cite{multi-probing}. The whole lattice sites are now divided into  $N_{mp}$ groups,
then we can produce $12$ ($12 = N_{c}\times N_{d}$) sources  $S(\{x_{i}\},c,d)=S(\{x_{i}\})\delta_{cc_{0}}\delta_{dd_{0}}$ for each multi-probing group, which only has
non-zero unity value on sites $\{x_{i}\}$ belong to the corresponding group. The element of $(\gamma_{5}D(0,x,x))_{cd}~, ~~x \in \{x_{i}\}$  can now be probed by
\begin{equation}\label{probing}
\begin{split}
   &  (\gamma_{5}D_{ov}(0,x,x))_{cd} \\
  \approx & \sum_{c_{0}d_{0}x_{0}c_{1}d_{1}x_{1}}
 \delta_{cc_{0}}\delta_{dd_{0}} \delta_{xx_{0}} (\gamma_{5}D_{ov}(0))_{\stackrel{c_{0}d_{0}x_{0}}{c_{1}d_{1}x_{1}}} S(\{x_{j}\})\delta_{cc_{1}}\delta_{dd_{1}} \\
  = & \sum_{c_{0}d_{0}x_{0}} \delta_{cc_{0}}\delta_{dd_{0}} \delta_{xx_{0}} (\gamma_{5}D_{ov}(0)S(\{x_{i}\},c,d))_{c_{0}d_{0}x_{0}}.
\end{split}
\end{equation}
Thus, only $12N_{mp}$ overlap operator multiplications with vector sources have to be performed with multi-probing method, which is to be compared with $12L_{x}L_{y}L_{z}L_{t}$ if calculating all-scale $q(x)$ directly, saving
a substantial amount of computations especially for large lattices.

In Tab.~\ref{tab:cnfg} we present the three ensembles of SU(3) pure gauge configurations in our study which are generated with periodic boundary using Iwasaki action. All have the same lattice size of $N_{s}=16$, $N_{t} = 32$, and we use Wilson flow with $w_{0} = 0.1670(10)\textrm{fm}$ to set the scale of lattice spacing $a$~\cite{wilsonflow1, wilsonflow2, wilsonflow3}. In the calculations of topological charge density, we choose the interval $\delta r = 4\sqrt{2}a $ to approximate the all-scale topological density $q(x)$, which ensures that almost all configurations have $\delta Q < 0.1 $, in which $\delta Q$ is the deviation of $Q=\displaystyle\sum_{x}q_{mp}(x)$ from an integer, since the massless overlap operator $D_{ov}(0)$ should satisfy Atiyah-Singer index theorem. Therefore only $4^{4}\times2\times3\times4=6144$ times of matrix multiplication to vector are needed, as for calculating all-scale $q(x)$ directly, $ 16^{3}\times 32 \times 3 \times 4 = 1572864$ times will be needed. In column $5$ and $6$ we list the percentage of the configurations that satisfy
 the condition $\delta Q $ below some arbitrary cut values, $0.08$ and $0.05$ respectively, and
 the obtained values for $\chi^{\frac{1}{4}}(0)$ using multi-probing method are tabulated in last column.
 It is noted that these values are compatible with phenomenological estimates $\chi^{YM} \approx(180\textrm{MeV})^4$~\cite{WittenVeneziano2, cqq1}.

\end{multicols}

\begin{table}[!h]
\begin{center}
\tabcaption{Three $16^3\times32$ ensembles with periodic boundary we used in this paper , the Iwasaki gauge action is used.}
 \begin{tabular} {|c|c|c|c|c|c|c|}
     \hline
     $\beta$ & $a[\textrm{fm}]$ & $(Ls/a)^3 \times (Lt/a)$& $N_{cnfg}$ & r($\delta Q <0.08$)[\%] & r($\delta Q$ <0.05)[\%] & $\chi^{\frac{1}{4}}(0) [\textrm{MeV}] $ \\
     \hline
     9.03 & 0.1161(10) & $16^3\times32$ & 200 & 95.5 & 72.5 & 190.7(61) \\
     \hline
     9.45 & 0.0963(9) & $16^3\times32$ & 200 & 97.5 & 77.5 & $188.4(60)$\\
     \hline
     10.02 & 0.0769(9) & $16^3\times32$ & 200 & 96.5 & 82.5 & 184.2(68) \\
     \hline
   \end{tabular}\label{tab:cnfg}
   \end{center}
\end{table}

\begin{figure}[!htb]
\centering
\subfigure[]{
\includegraphics[width=0.48\textwidth,height=0.28\textwidth]{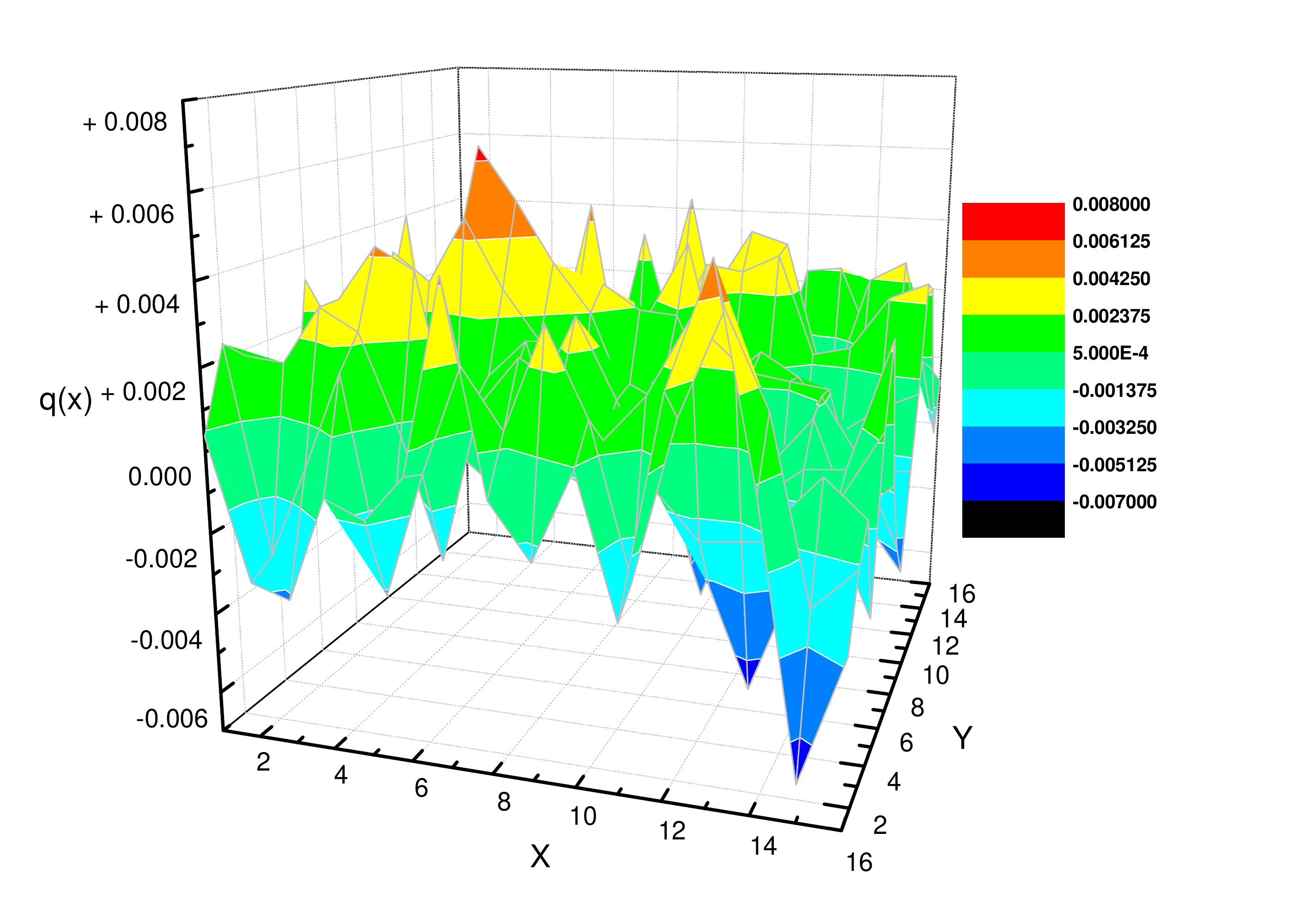}
\label{2dqx}
}
\subfigure[]{
\includegraphics[width=0.48\textwidth,height=0.28\textwidth]{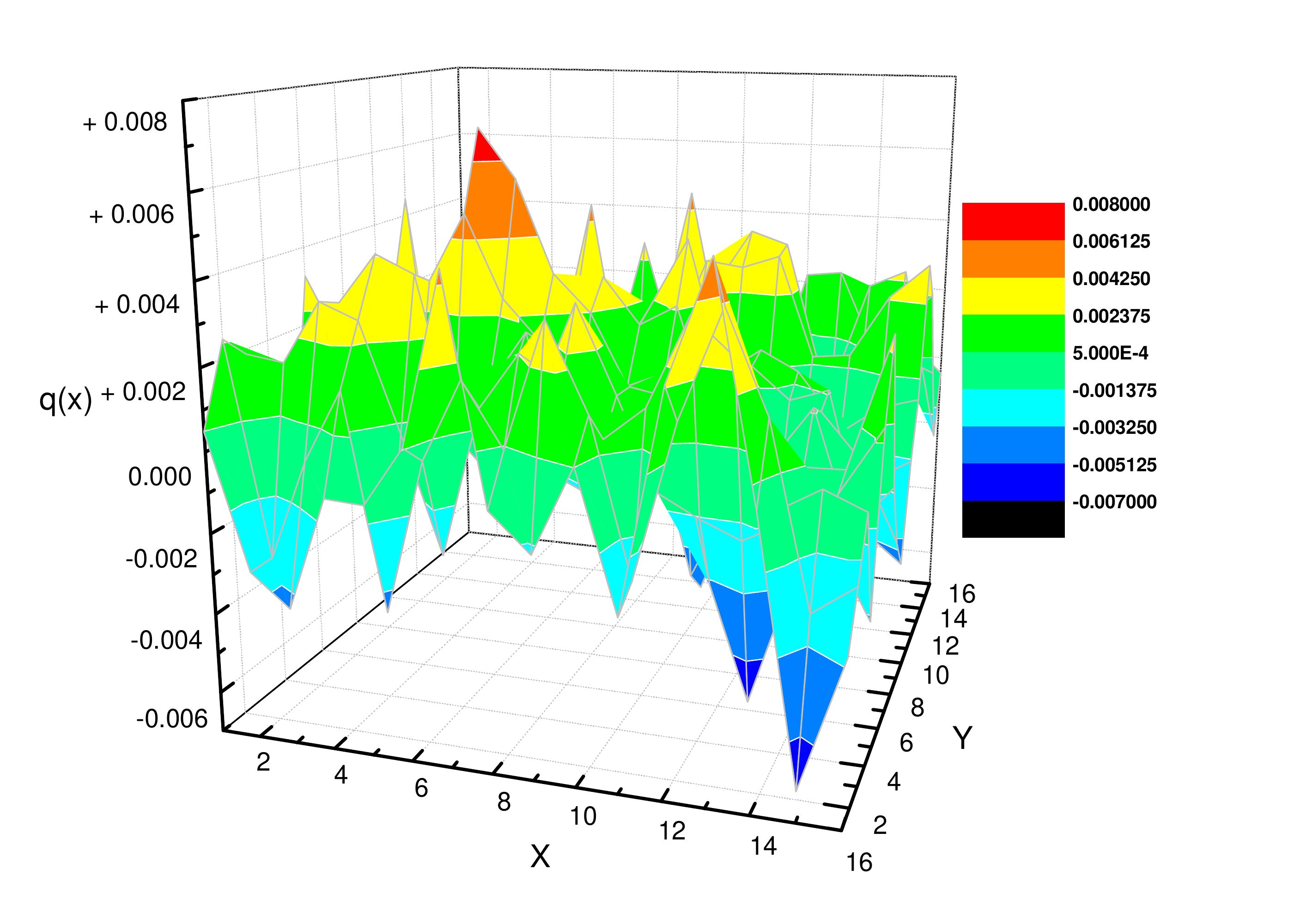}
\label{2dmpqx}
}
\caption{ Comparison of $q(x)$ on a $2D$ surface of one configuration with $Q=0$ of ensemble $a=0.0963$.
Left: all-scale $q(x)$. Right: $q(x)$ computed by multi-probing with $\delta r = 4\sqrt{2}a$. }
\label{fig:2dsketch}
\end{figure}

\begin{multicols}{2}
In Fig.~\ref{fig:2dsketch}, we present topological charge density computed both by all-scale $q(x)$ directly and multi-probing method with $\delta r = 4\sqrt{2}a$ on a $2D$ surface, Fig.~\ref{2dmpqx} only differs a little bit from Fig.~\ref{2dqx}, the features of $q(x)$ such as peaks, valleys and ridges are consistent, more detailed analyses of the structure of $q(x)$ will be discussed in Sect.~\ref{cluster:qx}.

\end{multicols}

\begin{figure}[!htb]
\centering
\subfigure[$2D$]{
\includegraphics*[viewport= 0cm 1cm 28cm 18cm, width=0.40\textwidth,height=0.25\textwidth]{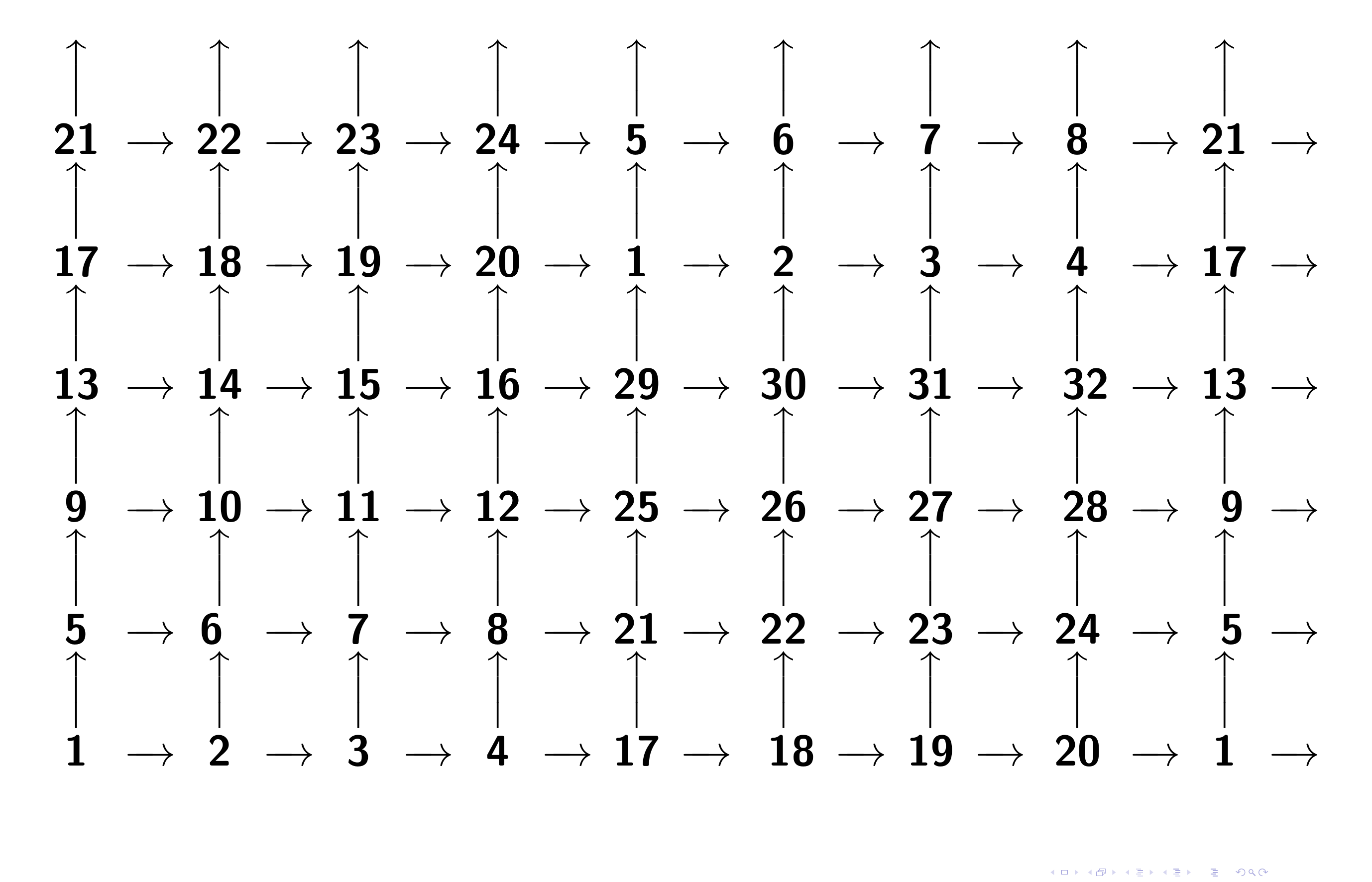}
\label{fig:2dprobing}}
\subfigure[$3D $]{
\includegraphics*[viewport= 0cm 1cm 11cm 11cm, width=0.40\textwidth,height=0.25\textwidth]{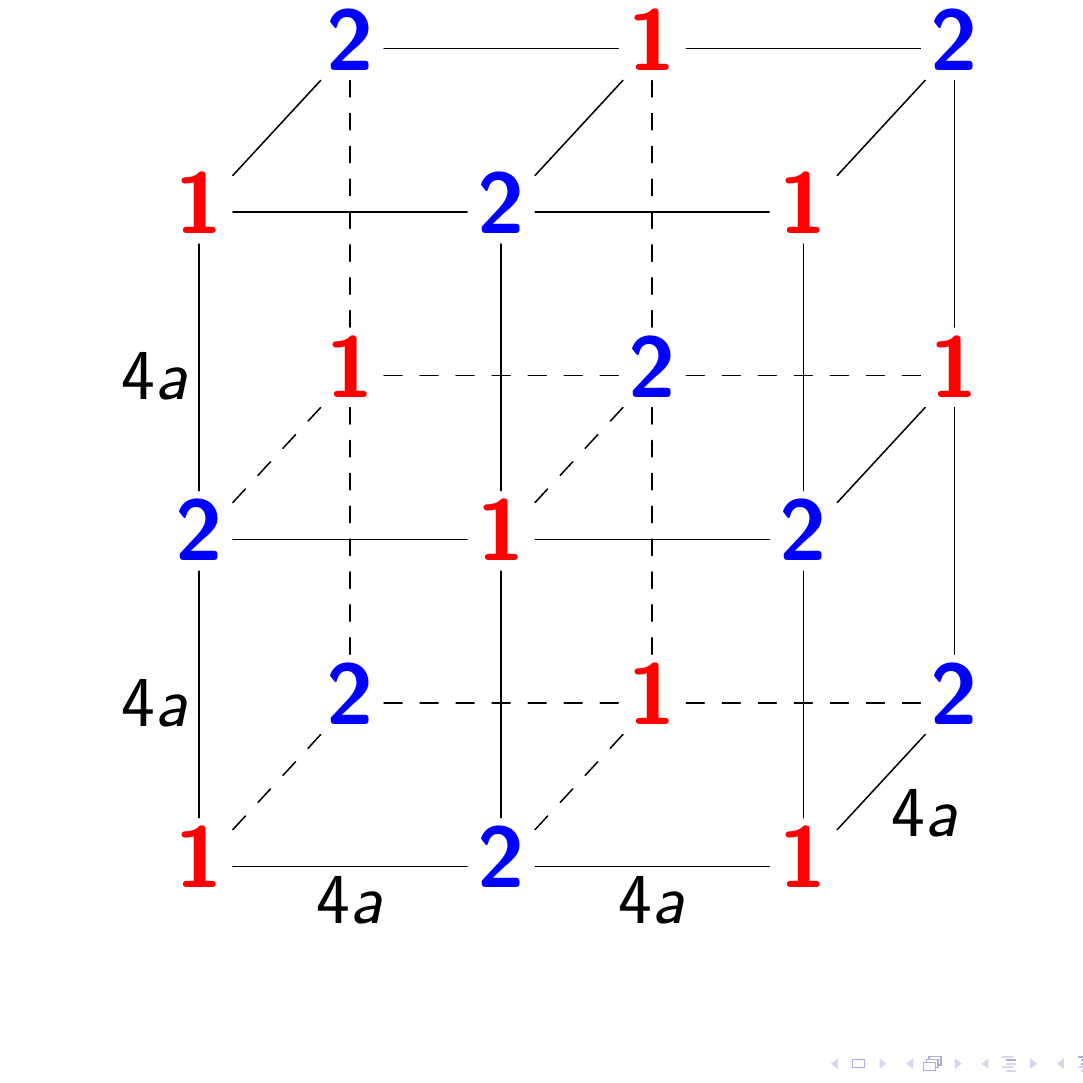}
\label{fig:3dprobing}}
\label{fig:multi_probing}
\caption{sketches for multi-probing scheme with interval $4 \sqrt{2}a$, sites with same number belong to the same multi-probing group. Arrows on the left panel are links with length equals lattice spacing $a$, and $4a$ on the right panel.}
\end{figure}

\begin{multicols}{2}
The scheme of multi-probing  with interval $\delta r = 4 \sqrt{2}a$ is presented on $2D$ surface in Fig.~\ref{fig:2dprobing}, sites with same number belong to the same multi-probing group, and $3D$ sketch Fig.~\ref{fig:3dprobing} shows that the diagonal points, which are on the surfaces of cube with side length equals $4a$ in the lattice, belong to the same multi-probing group. As to the scheme of multi-probing with interval $\delta r = 4a$, vertices of cube with side length equals $4a$ should belong to the same multi-probing group, in other words, sites with number $1$ and $2$ in Fig.~\ref{fig:3dprobing} belong to the same group now, then the number of groups will be reduced by half. Therefore the schemes of multi-probing  with interval $\delta r = 4 \sqrt{2}a$ and $\delta r = 4a$ are correlated strongly, same relationship exists between schemes with interval $\delta r = n \sqrt{2}a$ and $\delta r = n a$ universally.

\end{multicols}

\begin{multicols}{2}

\section{Analysing the topological charge density $q(x)$}\label{fluctuation}

Noting that topological susceptibility can be computed from topological charge individually, and the topological charge $Q$ should be an integer value, which can be used to check the degree of approximation. Although the multi-probing method seems to approximate $Q$ rather well as indicated in Tab.~\ref{tab:cnfg}, one is still concerned about the performance of topological charge density and especially the slope of the susceptibility.
Thus, we need to find  some other parameters for testing the validity of the approximation. From Eq.~\eqref{chiprimeLQCD} we may compare the tow-point correlator $C_{qq}(r)$ from multi-probing and all-scale $q(x)$. Other tools proposed in Ref.~\cite{qx:structure3} will be utilized to analyze the vacuum structure of topological charge density, results will also be compared with that obtained from all-scale $q(x)$.

\subsection{$C_{qq}(r)$ from multi-probing method }

The  reflection positivity~\cite{reflectionpositivity} requires that two-point correlator of topological charge density should be negative~\cite{cqq1, cqq2, cqq3}:
\begin{equation}\label{cqq:negative}
    C_{qq}(r) \leq 0 ~~for ~~r > 0.
\end{equation}
Together with massless overlap operator $D_{ov}(0)$ has a finite range, $C_{qq}(r)$ with overlap fermions will have a positive core, which has been studied extensively already.

\begin{figure}[!htb]
\centering
\hspace{7pt}
\subfigure[]{
\begin{overpic}[width=0.43\textwidth,height=0.28\textwidth]{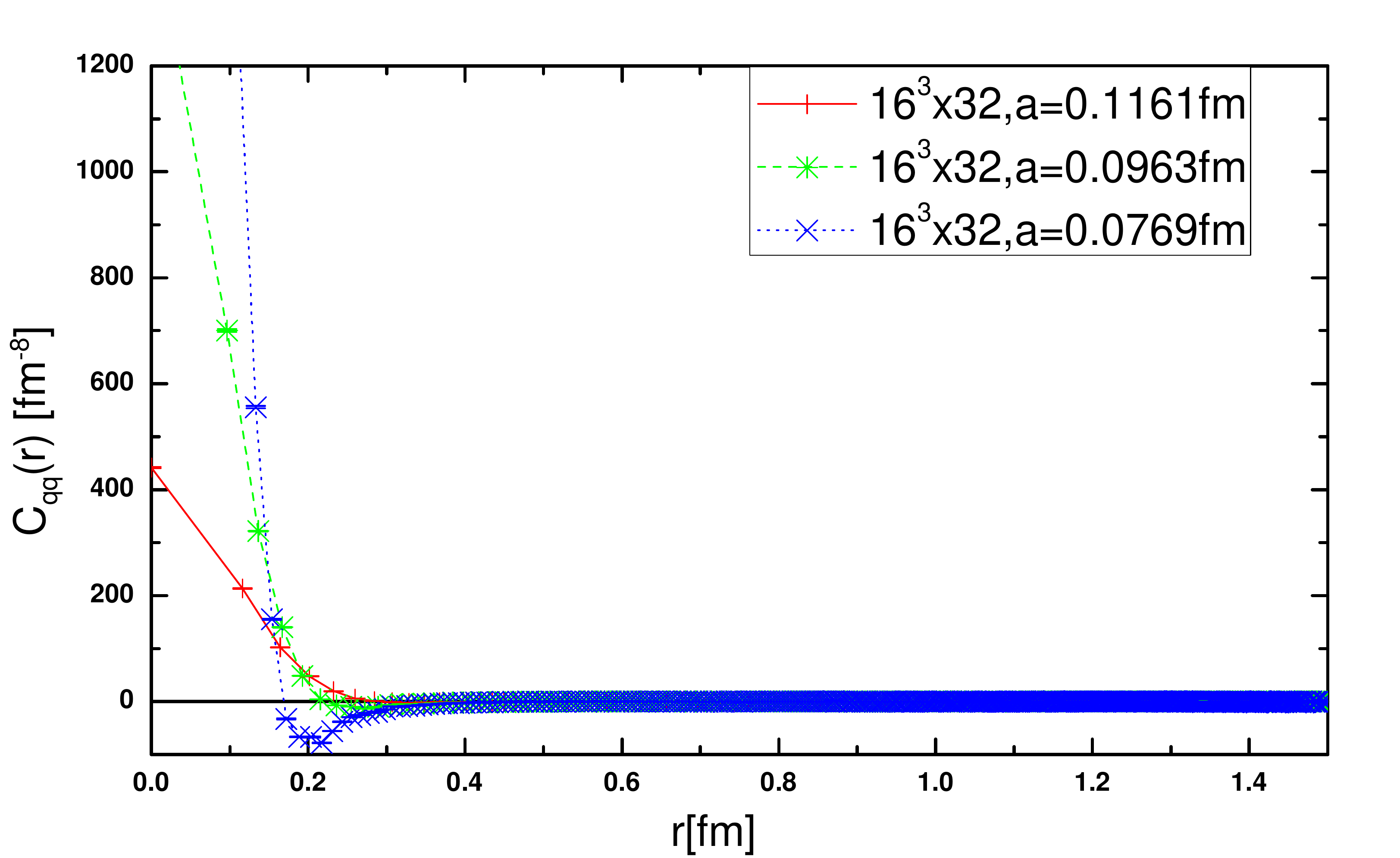}
\put(30,14){\includegraphics[width=0.25\textwidth,height=0.13\textwidth]{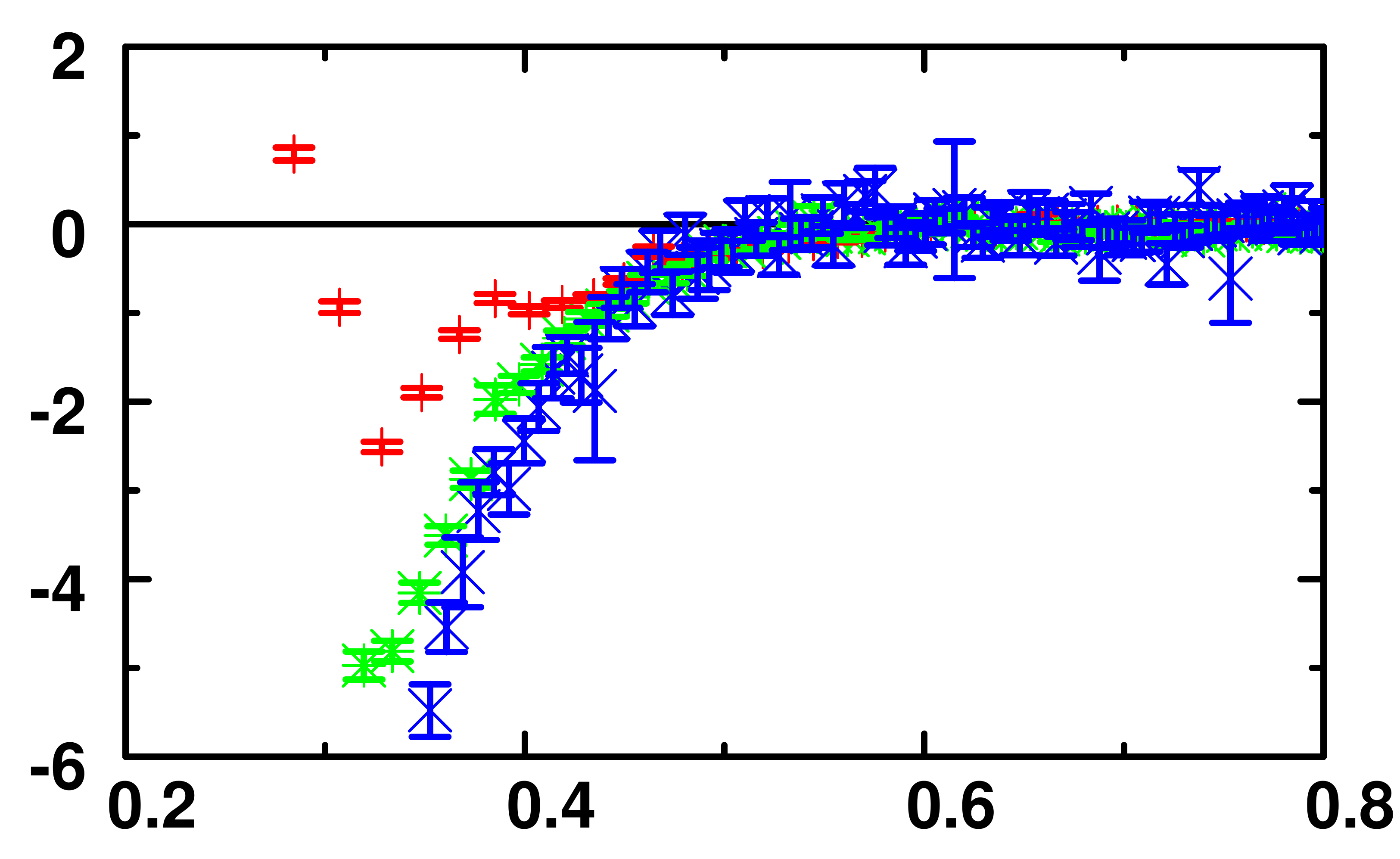}}
\end{overpic}
\label{fig:cqq1}
}
\subfigure[]{
\includegraphics[width=0.43\textwidth,height=0.28\textwidth]{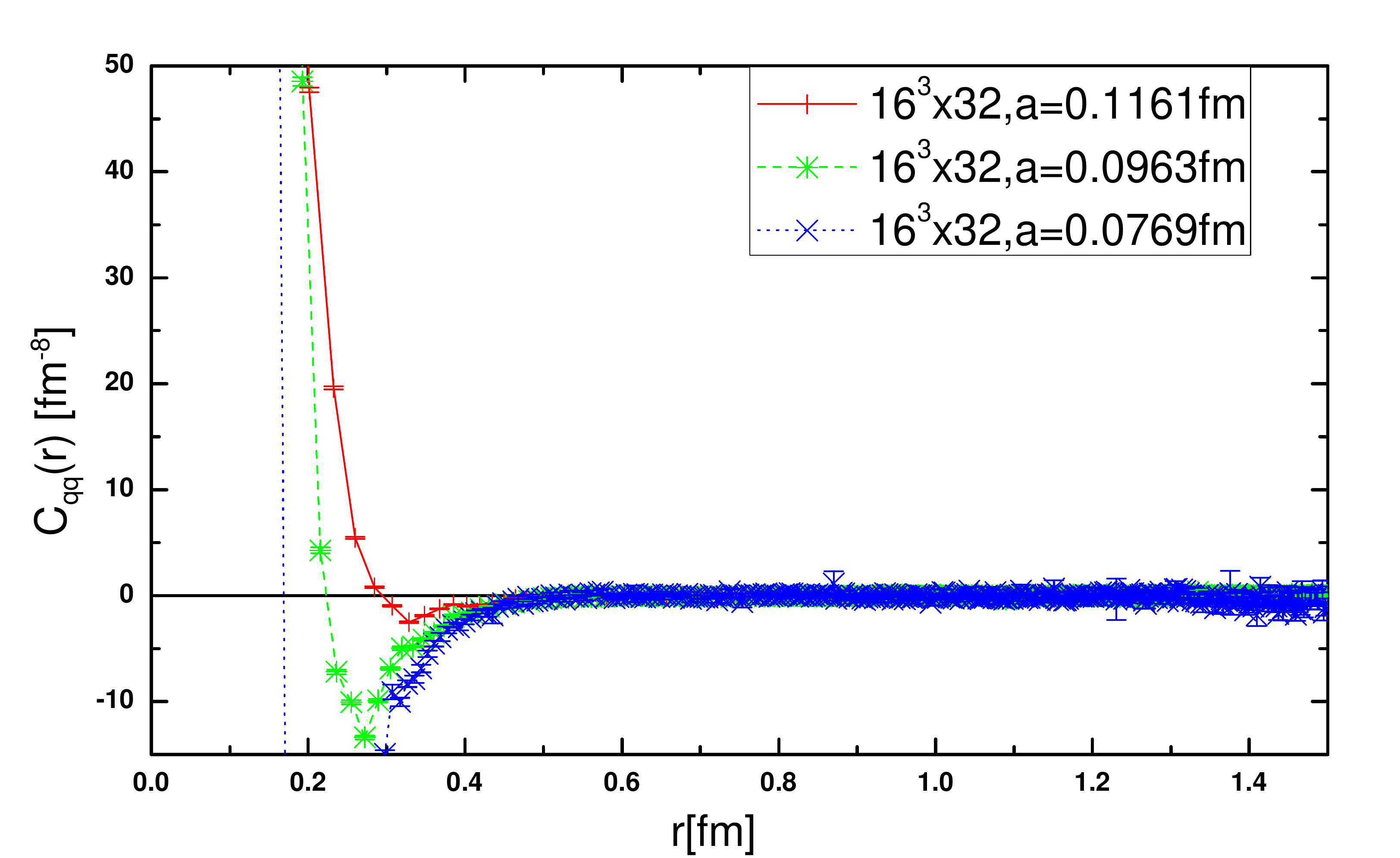}
\label{fig:cqq2}
}
\subfigure[Fig.14(a) in Ref.~\cite{qx:structure3}]{
\includegraphics*[viewport=4.5cm 15cm 17.3cm 23.5cm, width=0.43\textwidth,height=0.28\textwidth]{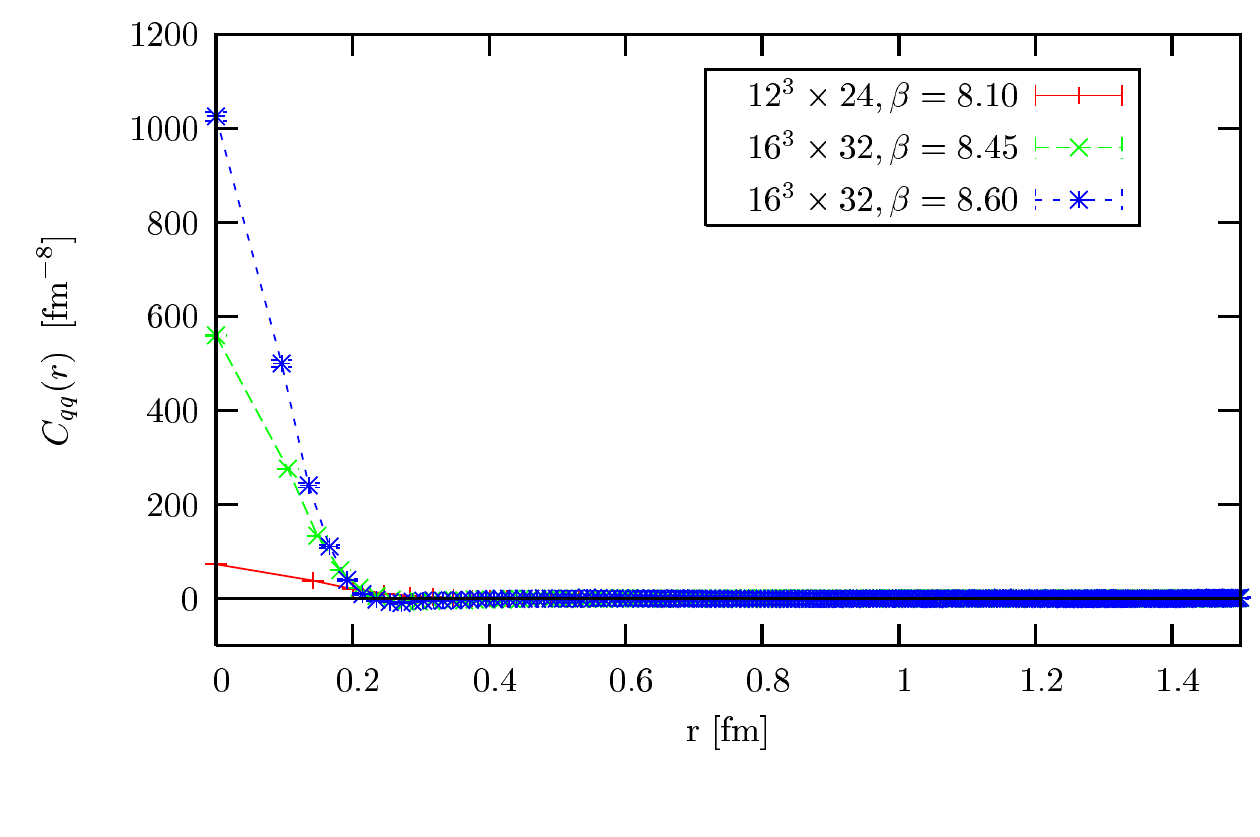}
\label{fig:cqq3}
}
\subfigure[Fig.14(b) in Ref.~\cite{qx:structure3}]{
\includegraphics*[viewport=4.5cm 15cm 17.3cm 23.5cm, width=0.43\textwidth,height=0.28\textwidth]{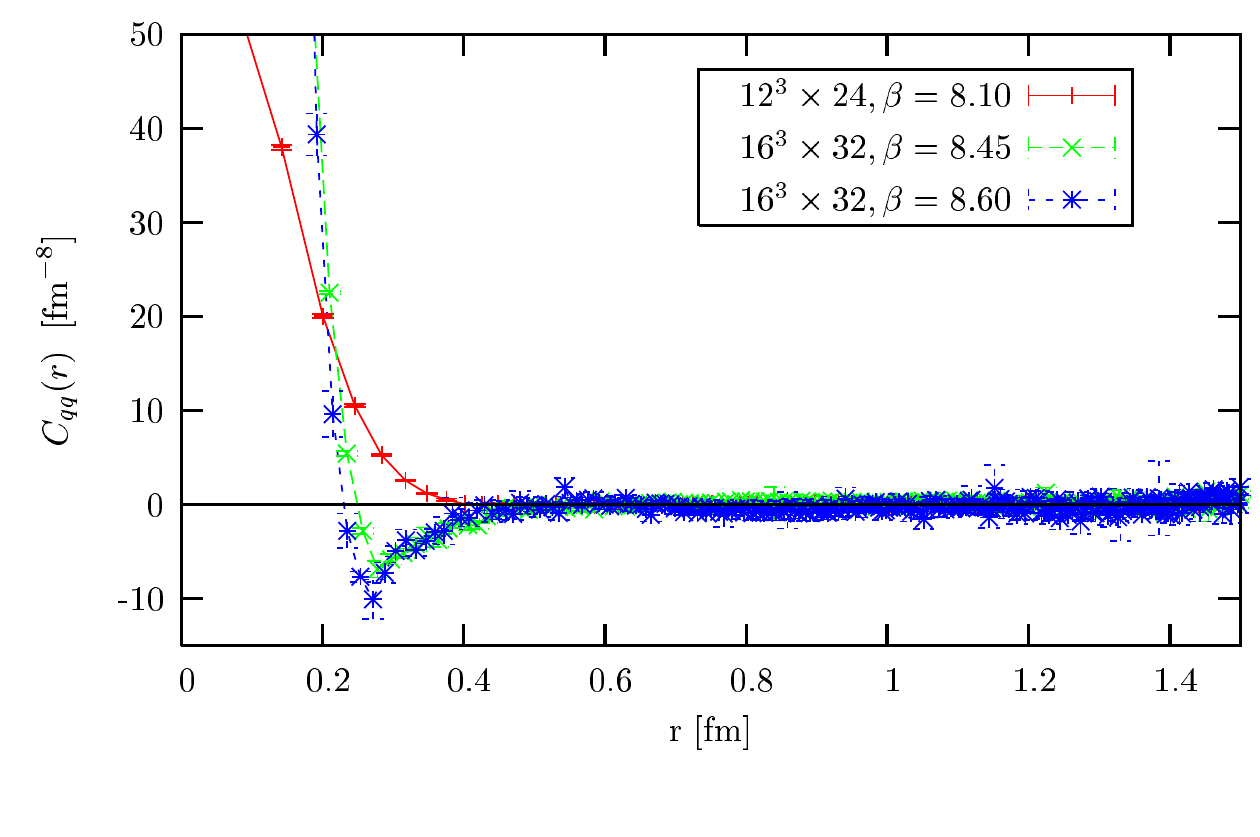}
\label{fig:cqq4}
}
\label{fig:cqq}
\caption{ Comparison of $C_{qq}(x)$. Top row: Results of multi-probing. Bottom row: Results of all-scale $q(x)$ from Fig.14 in~\cite{qx:structure3}. Noting that the scale is much smaller in the right side, and the upper left figure is embedded with partial enlarged detail. }
\end{figure}

In Figs.~\ref{fig:cqq1},~\ref{fig:cqq2} the two-point correlator from our work displays positive core and negative tails, just as all-scale $q(x)$ does in Figs.~\ref{fig:cqq3},~\ref{fig:cqq4} ( results of Fig.14 in Ref.~\cite{qx:structure3}). The details showed in Figs.~\ref{fig:cqq2},~\ref{fig:cqq4} demonstrate that the ranges of positive cores  are  consistent with $\beta=9.45~(a\approx 0.1fm)$ in Fig.~\ref{fig:cqq2} and of $\beta=8.45, 8.60 (a\approx 0.1fm)$ in Fig.~\ref{fig:cqq4} with the same lattice size $16^{3}\times 32$. From Fig.~\ref{fig:cqq2}, ensemble with different lattice spacing $a$ has different positive core in physical unit, but almost the same in lattice unit, which means it is independent of $\beta$ in lattice unit, this effect comes from the fact that the finite range of overlap operator $D_{ov}(0)$ has little dependence on $\beta$. In other words, as lattice spacing $a$ decreases, the positive core of $C_{qq}(x)$ will disappear in physical unit, when $a \rightarrow 0$, only the contact term $C_{qq}(0) > 0 $,
and the reflection positivity condition Eq.~\eqref{cqq:negative} will be restored in the continuum limit.

\end{multicols}

\begin{multicols}{2}

Inspired by the instanton model, a functional form of the two-point correlator $C_{qq}(x)$ in the negative region can be used to extract pseudoscalar particle mass~\cite{Cqq_mass}:
\begin{equation}\label{cqq:mass1}
  \langle q(x)q(y)\rangle = \frac{m}{4\pi^{2}r} K_{1}(mr),~~r=\mid x-y \mid ,
\end{equation}
where $K_{1}(z)$ is the modified Bessel function which has asymptotic form as
\begin{equation}\label{K1func}
  K_1(z) ~\underset{{\rm large}~z}{\sim} ~e^{-z} ~ \sqrt{\frac{\pi}{2z}}~ \left[1 + \frac{3}{8z}  \right ].
\end{equation}
 We can use Eqs.~\eqref{cqq:mass1}~\eqref{K1func} to fit the data of $C_{qq}(r)$. In the fitting procedure the amplitude and mass are treated as free parameters and then we can extract mass of pseudoscalar glueball in quenched QCD, since only glueballs exist in pure gauge theory. The fitting range is determined by the effective mass $m_{eff}(r)$ which is defined by
 \begin{equation}\label{effective_mass}
  \frac{C_{qq}(r+\Delta r)}{C_{qq}(r)} = \frac{r}{r+\Delta r} \frac{K_{1}(m_{eff}(r+\Delta r))}{K_{1}(m_{eff}r)},
\end{equation}
where we set $\Delta r = 0.5a$ and $C_{qq}(r)$ will be averaged over $[r-\frac{\Delta r}{2},r+\frac{\Delta r}{2}]$.

\begin{figure}[!htb]
\centering
\subfigure[a=0.0769fm]{
\includegraphics[width=0.43\textwidth,height=0.24\textwidth]{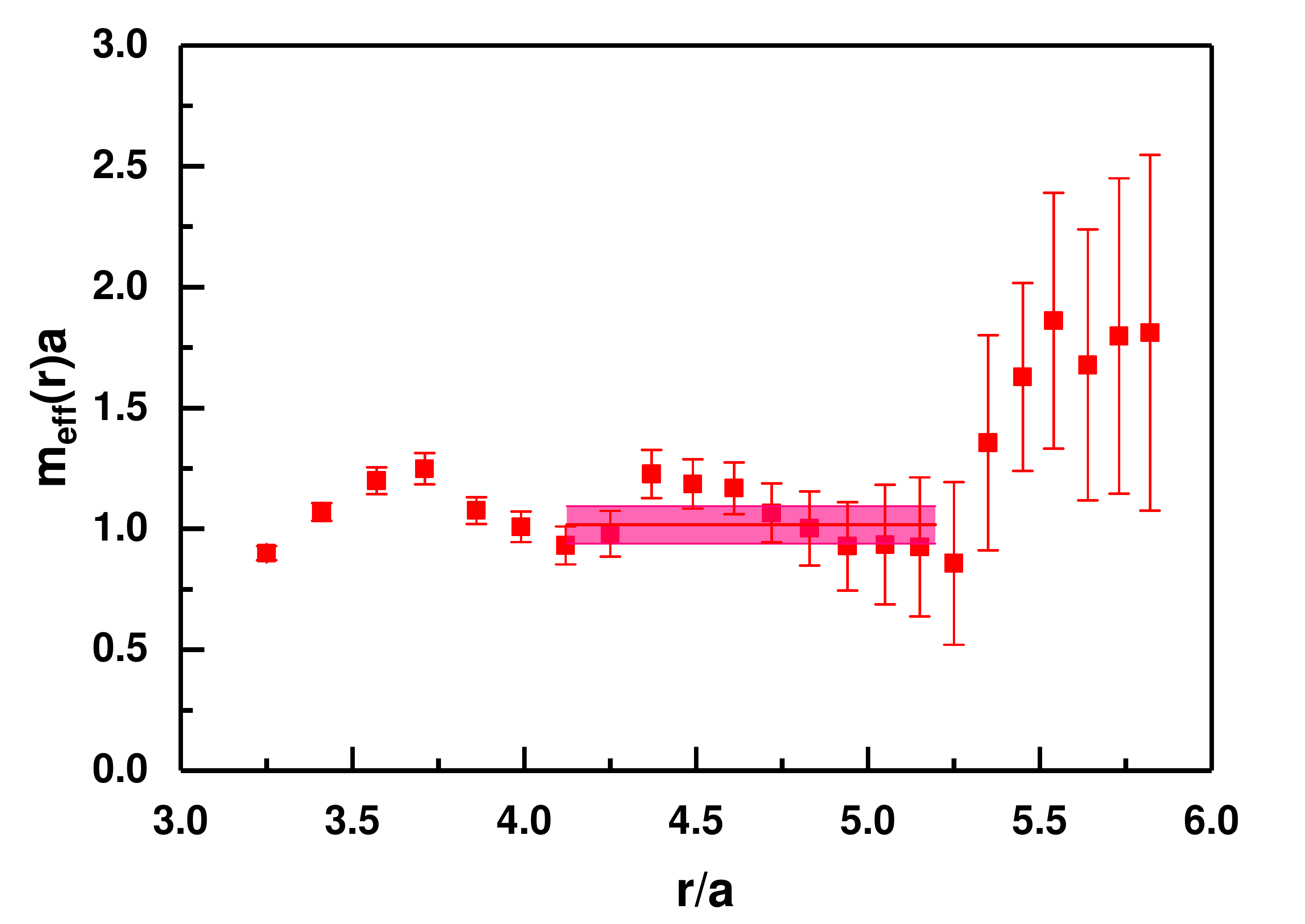}
\label{fig:mass1}
}
\subfigure[a=0.0963fm]{
\includegraphics[width=0.43\textwidth,height=0.24\textwidth]{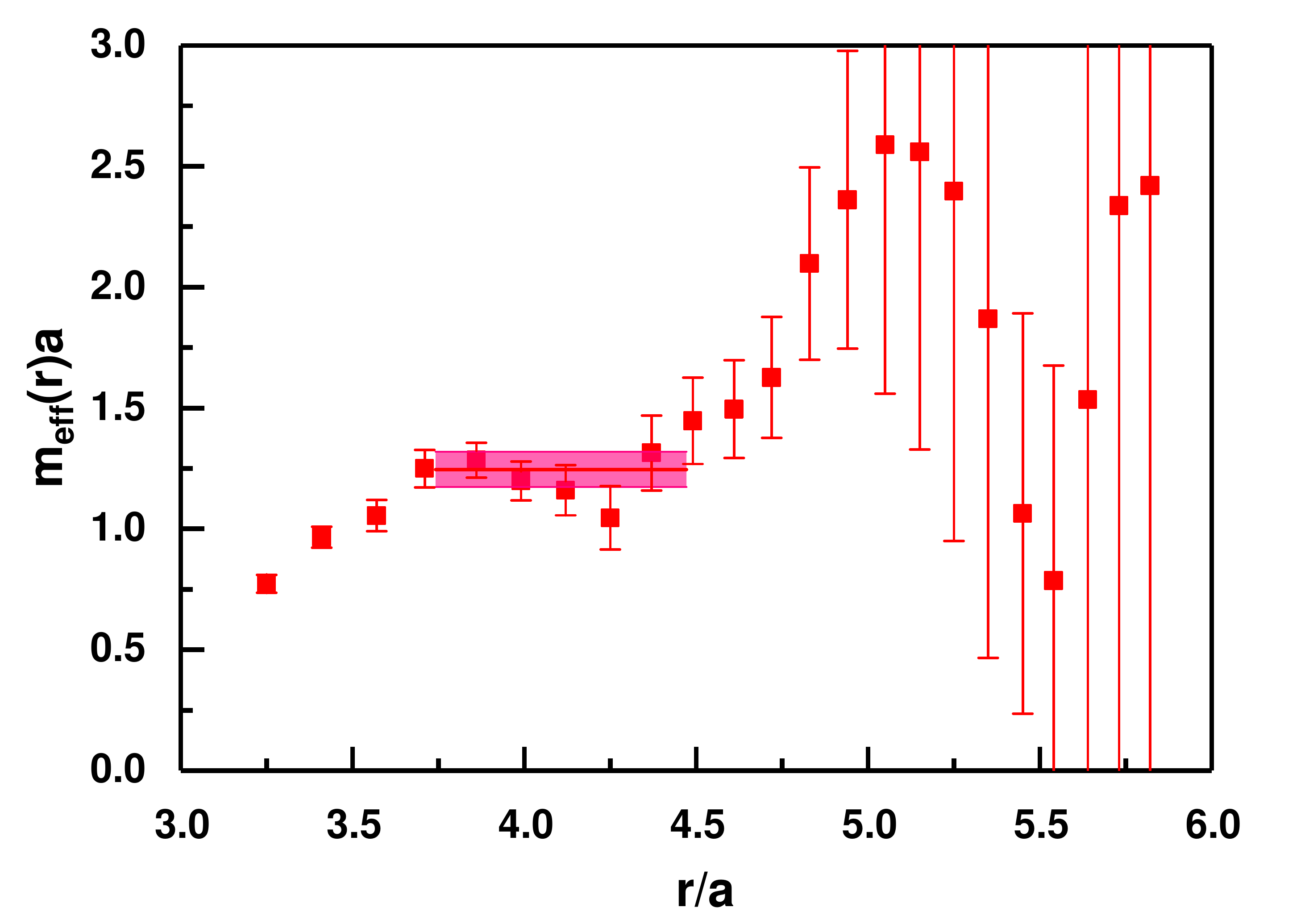}
\label{fig:mass2}
}
\caption{ Plateaus of effective pseudoscalar glueball mass,left panel for ensemble $a=0.0769\textrm{fm}$, right panel for ensemble $a=0.0963\textrm{fm}$. The color bands shows the fitting mass with errors and also the fitting range. }
\label{fig:effective_mass}
\end{figure}

In Fig.~\ref{fig:effective_mass} we present the effective mass  plateaus of $2$ ensembles in Tab.~\ref{tab:cnfg} with errors computed by Jackknife method, except ensemble $a=0.1161\textrm{fm}$, whose lattice spacing may be too coarse. The starting and ending points of fitting range would be varied in $[3.25a,5.25a]$ for $a=0.0769\textrm{fm}$ ensemble and $[3.5a,4.5a]$ for $a=0.0963\textrm{fm}$ ensemble respectively, they are the plateaus shown in Fig.~\ref{fig:effective_mass}, and fitting range is set to be greater than or equal to $\Delta r = 0.5a$, then the final fitting range is decided by $\chi^{2}/d.o.f.$ that nearest to $1.0$~\cite{fitting01}.

\end{multicols}
\begin{multicols}{2}
Tab.~\ref{tab:gluemass} presents the fitted results for the masses for ensembles $a=0.0963,~0.0769\textrm{fm}$, both
 of which are compatible with the value $2560(35)\textrm{Mev}$ from Ref.~\cite{glueballmass}, though the fitting range is determined somewhat subjectively. It is also seen that the mass has little dependence on the lattice spacing $a$.
\end{multicols}

\begin{table}[!htb]
\caption{Fitting mass of pseudoscalar glueball mass using asymptotic formula Eq.~\eqref{K1func}.}
\begin{center}
 \begin{tabular}{|c|c|c|c|}
   \hline
   $a[\textrm{fm}]$ & $ma$ & $m [\textrm{MeV}]$ & $\chi^{2}/d.o.f.$ \\
   \hline
   0.0769(9) & 1.017(78) & 2609(230) & 1.12\\
   \hline
   0.0963(9) &  1.246(73) & 2553(173) & 0.94 \\
   \hline
 \end{tabular}\label{tab:gluemass}
\end{center}
\end{table}

\begin{multicols}{2}

\subsection{clusters of topological charge density from multi-probing method  }\label{cluster:qx}

We use cluster analysis proposed in Ref.~\cite{qx:structure3} to investigate the sign-coherent structure of topological charge density (the word "sign-coherent" will be omitted below for simplicity):
\begin{enumerate}
  \item The cluster on the lattice is composed of sites that link-connected of same sign of the topological charge density $q(x)$, $q_{cut}$ means that only sites with $|q| > q_{cut} $ are taken into account, and $q_{max}$ is the maximum value of $|q(x)|$.

  \item The connectivity is defined by
  \begin{equation}\label{cluster:connectivity}
    f(r_{max}) = \frac{\sum_{x,y}\langle\sum_{c}\Theta_{c}(x)\Theta_{c}(y)\rangle\delta(r_{max}-|x-y|)}{\sum_{x,y}\delta(r_{max}-|x-y|)},
  \end{equation}
  $r_{max}=\sqrt{{L_x}^{2}+{L_y}^{2}+{L_z}^{2}+{L_t}^{2}}/2$ is the largest distance available on the periodic lattice, $\Theta_{c}(x)$ is a character function of the cluster c, namely $\Theta_{c}(x)=1~,~if~x\in c~,else~\Theta_{c}(x)=0$. The value $q_{cut} = q_{perc}$ for which $ f(r_{max}) = 0$ is called the onset of percolation. When $q_{cut} < q_{perc}, f(r_{max}) > 0$ indicating that global clusters exist.

  \item The Euclidean distance between the two largest clusters $c$ and $c^\prime$ is defined by :
  \begin{equation}\label{cluster:distance}
    d(c,c^\prime) = {\max_{x\in c}}({\min_{y\in c^\prime}}(|x-y|)).
  \end{equation}
  \item Random walker method is utilized to define the fractal dimension $d^{\ast}$ of a cluster. Starting from the site that has the maximum $q_{max}$ inside a cluster, and after $\tau$ steps of independent random walkers, the probability of returning to the starting point $P(\tau)=(2\pi \tau)^{-d^{\ast}/2}$. The walker jumps to linked site in the same cluster with equal probability  every step.
\end{enumerate}

Since the physics of vacuum is thought to be mostly affected by intense fields, and the super-long-distance might be relevant to the physics of the QCD vacuum such as Goldstone boson propagation~\cite{qx:structure2},
therefore we are concerned about the large topological charge density ($q_{cut}$ is large), at the same time global structures still survive, i.e. $f(r_{max}) > 0$.
\end{multicols}
\begin{figure}[!hbt]
\subfigure[]{
\includegraphics[width=0.5\textwidth,height=0.28\textwidth]{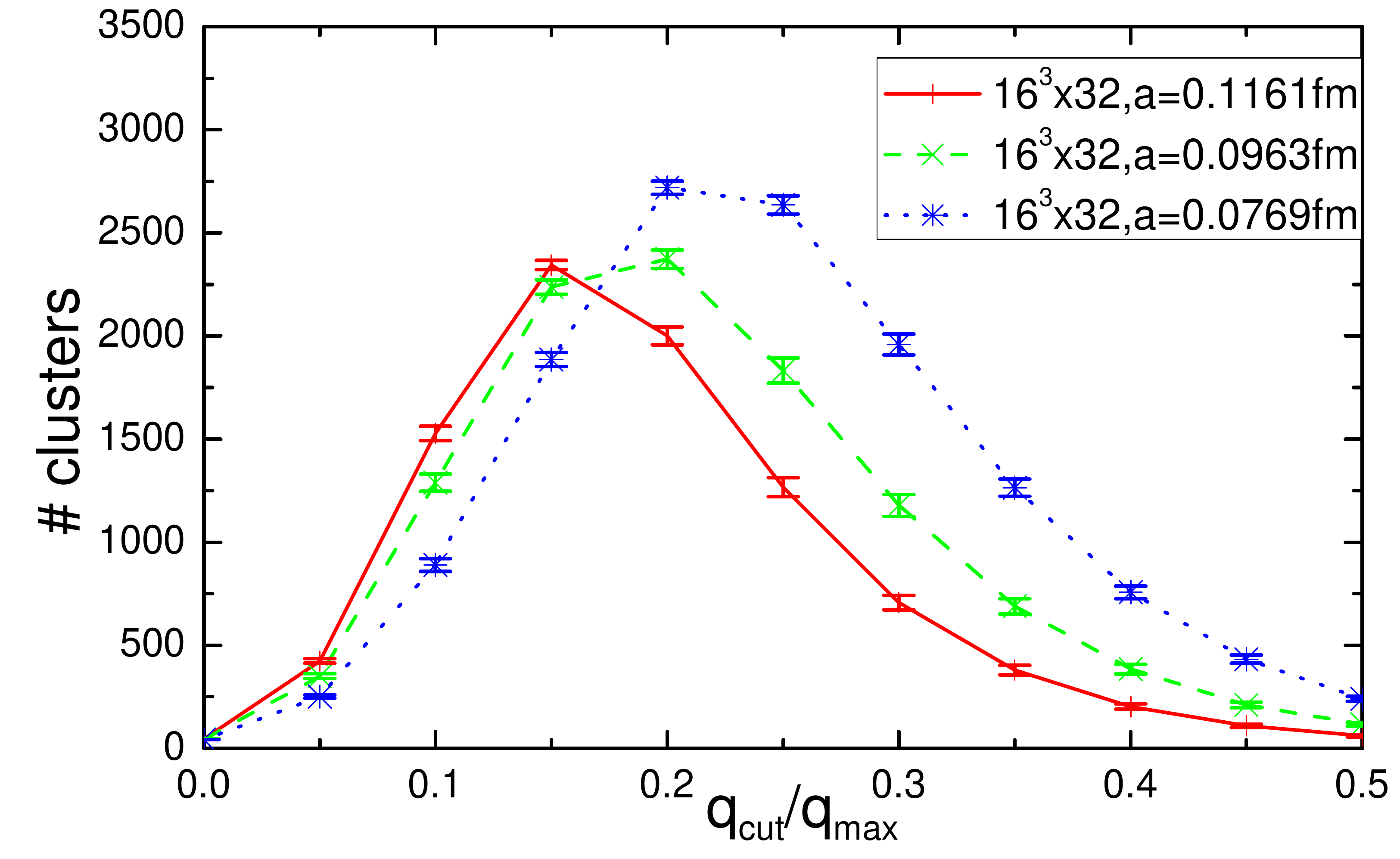}
\label{fig:clusters}
}
\subfigure[]{
\includegraphics*[viewport=4.5cm 15.8cm 17.8cm 23.3cm, width=0.5\textwidth,height=0.28\textwidth]{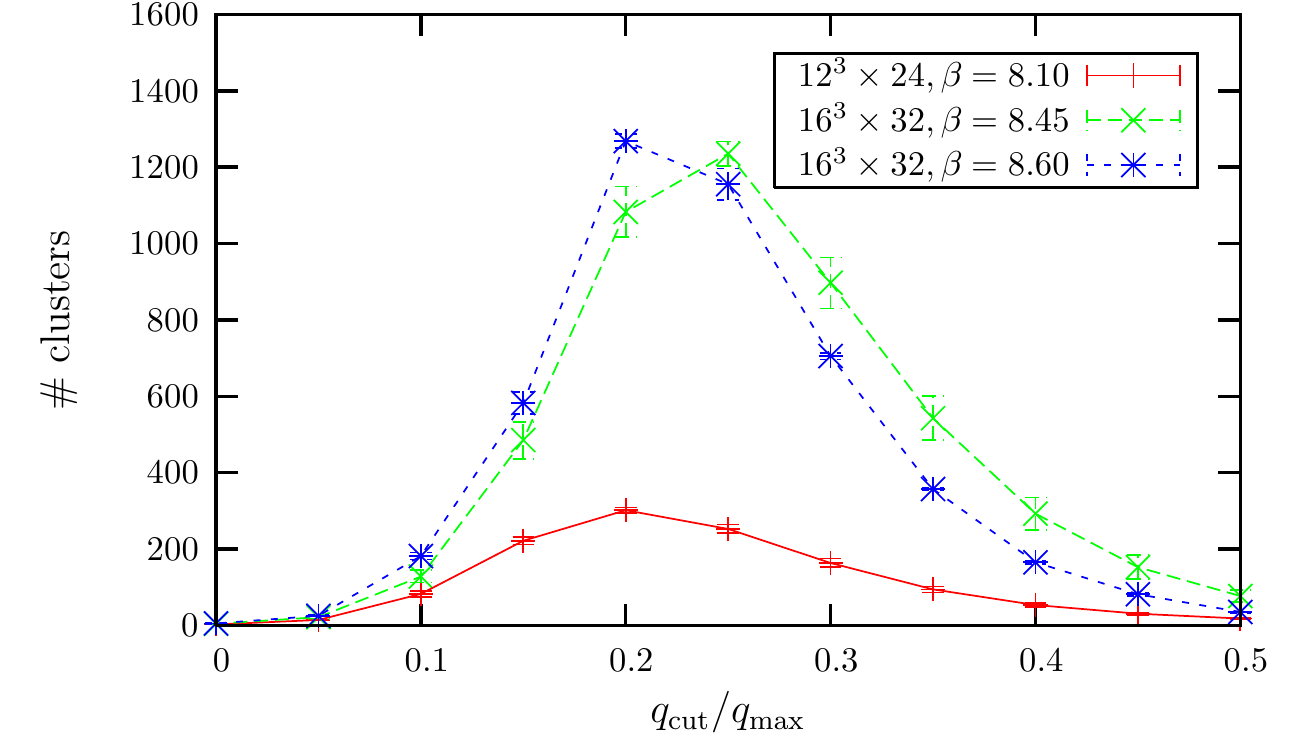}
\label{fig:refclusters}
}
\subfigure[]{
\includegraphics[width=0.5\textwidth,height=0.28\textwidth]{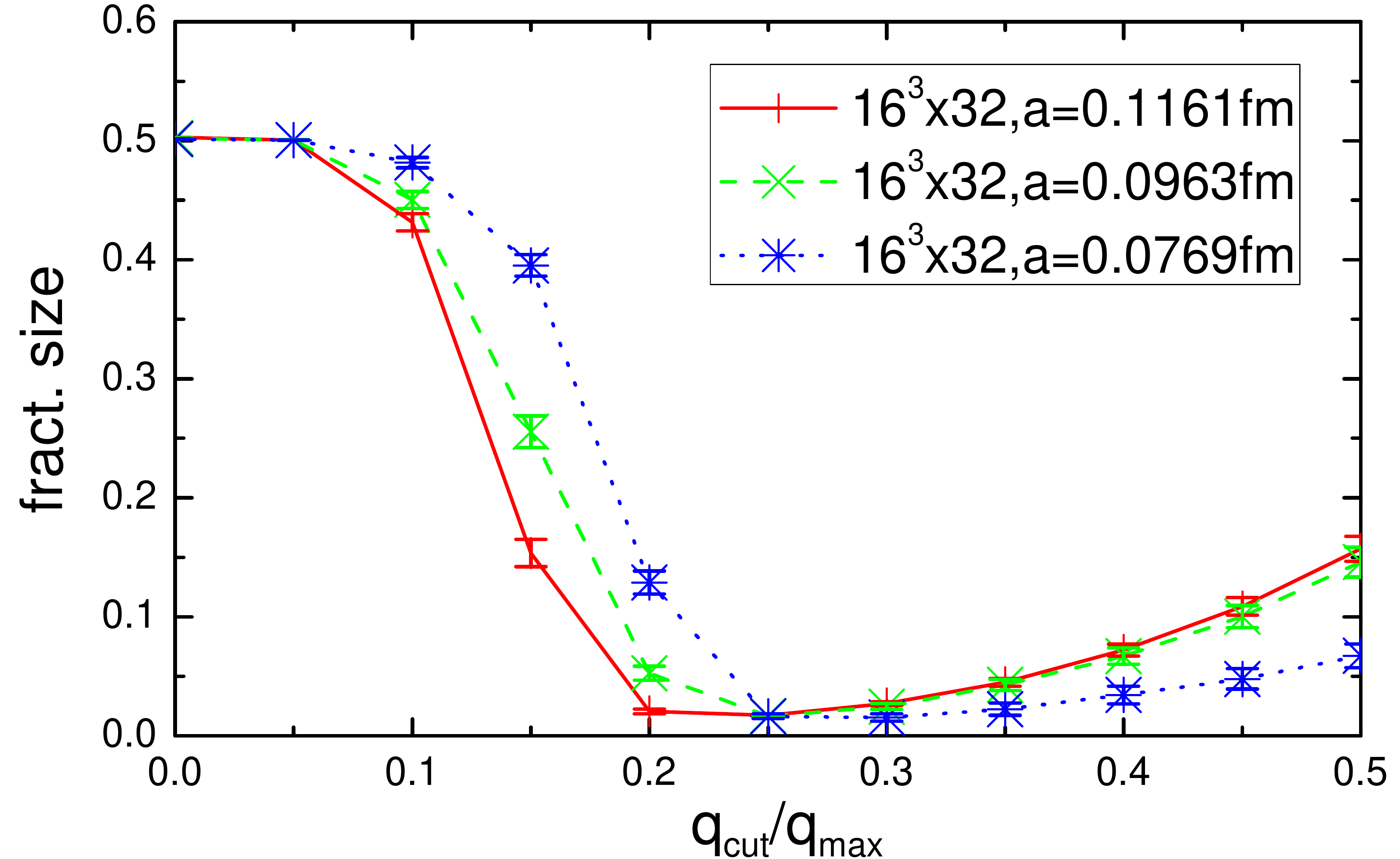}
\label{fig:frac}
}
\subfigure[]{
\includegraphics*[viewport=4.5cm 15.8cm 17.8cm 23.3cm, width=0.5\textwidth,height=0.28\textwidth]{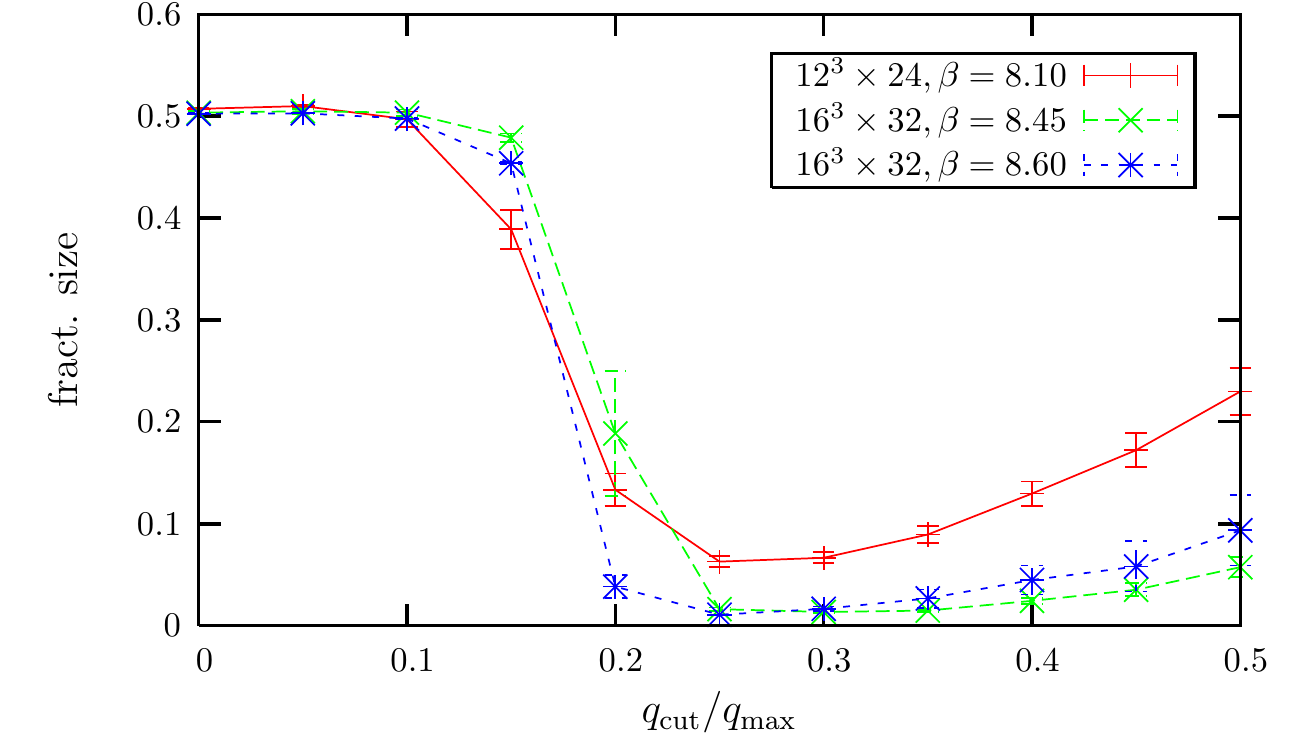}
\label{fig:reffrac}
}
\caption{ Left side from multi-probing method, right side from all-scale $q(x)$ in Ref.~\cite{qx:structure3}. 1st row: the total number of clusters. 2nd row: the size of the largest cluster relative to all clusters. }
\label{fig:qxstructure1}
\end{figure}

\begin{multicols}{2}
In Fig.~\ref{fig:qxstructure1}, the number of clusters, which we call cluster multiplicity, increases as $q_{cut}$ increases  till it reaches the maximum, then decreases; and the fractional size of the largest cluster, which denote the size of the largest cluster relative to total volume occupied by all clusters, reaches the minimum near the maximum point of cluster multiplicity  , then rises. The trend of fractional size of the largest cluster is just opposite to cluster multiplicity. At $ q_{cut} = 0 $ the two largest clusters with opposite sign both have fractional size of the largest cluster $\approx 0.5 $. Except the two largest clusters, the dozens of clusters left are fragments only with volume of $1 ~\textrm{or}~ 2 $, and their total volume compared with the lattice volume $16^{3}\times 32$ are so small that can be ignored.
\end{multicols}
\begin{figure}[!hbt]
\subfigure[]{
\includegraphics[width=0.5\textwidth,height=0.28\textwidth]{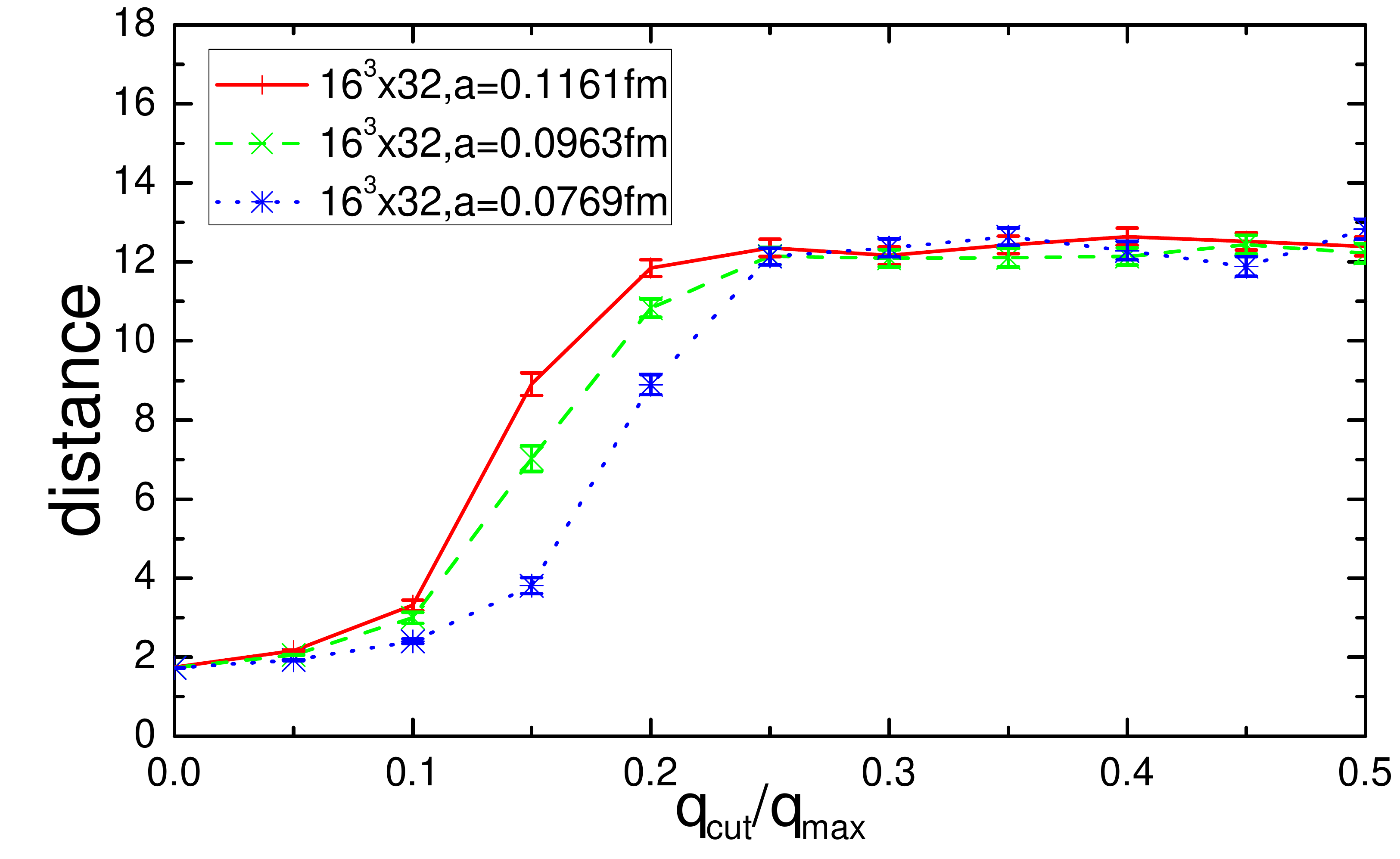}
\label{fig:distance}
}
\subfigure[]{
\includegraphics*[viewport=4.5cm 15.8cm 17.5cm 23.3cm, width=0.5\textwidth,height=0.28\textwidth]{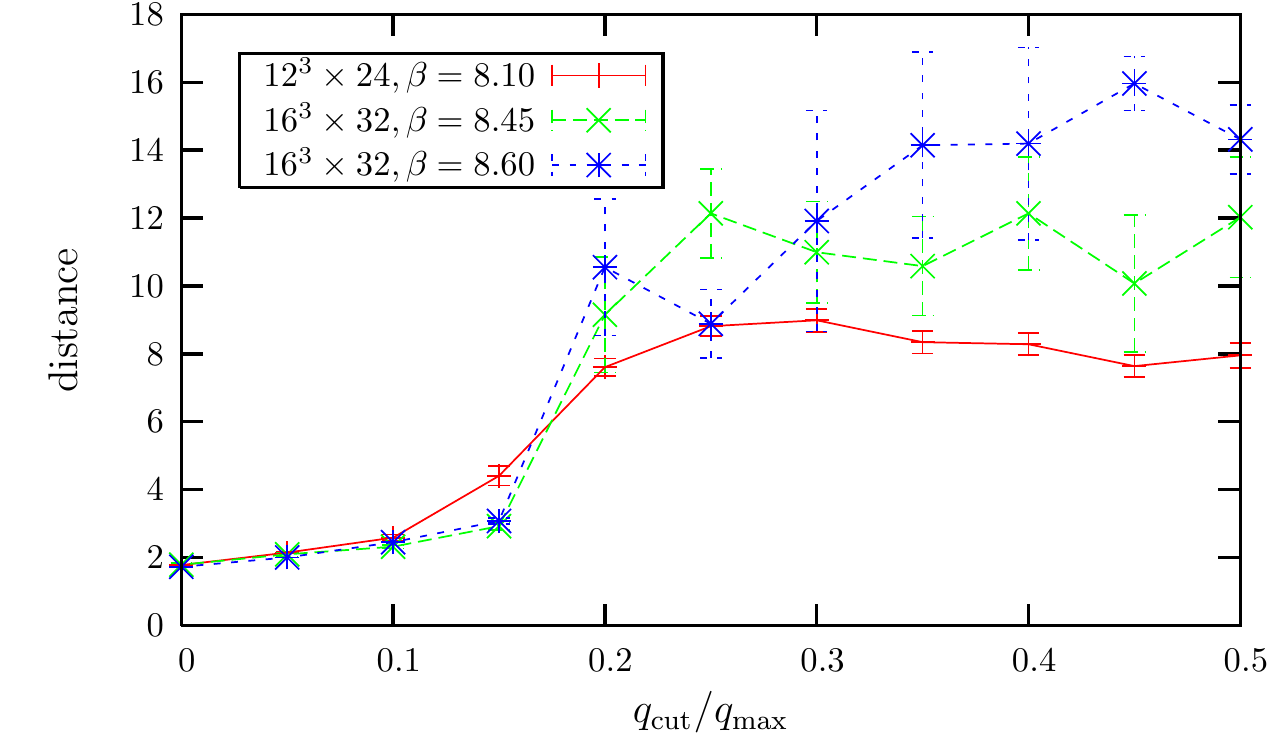}
\label{fig:refdistance}
}
\subfigure[]{
\includegraphics[width=0.5\textwidth,height=0.28\textwidth]{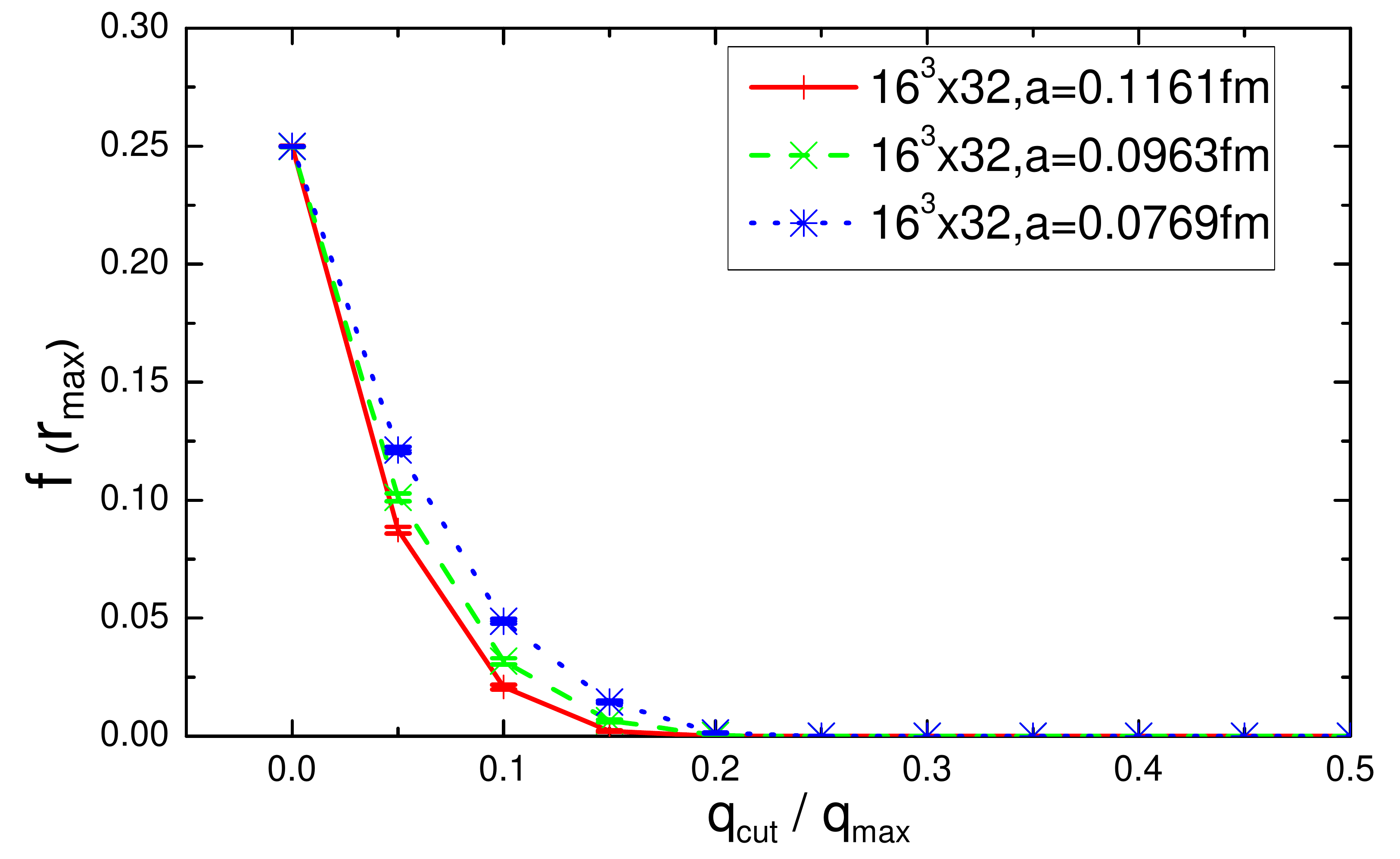}
\label{fig:connectivity}
}
\subfigure[]{
\includegraphics*[viewport=4.8cm 15.8cm 17.5cm 23.3cm, width=0.5\textwidth,height=0.28\textwidth]{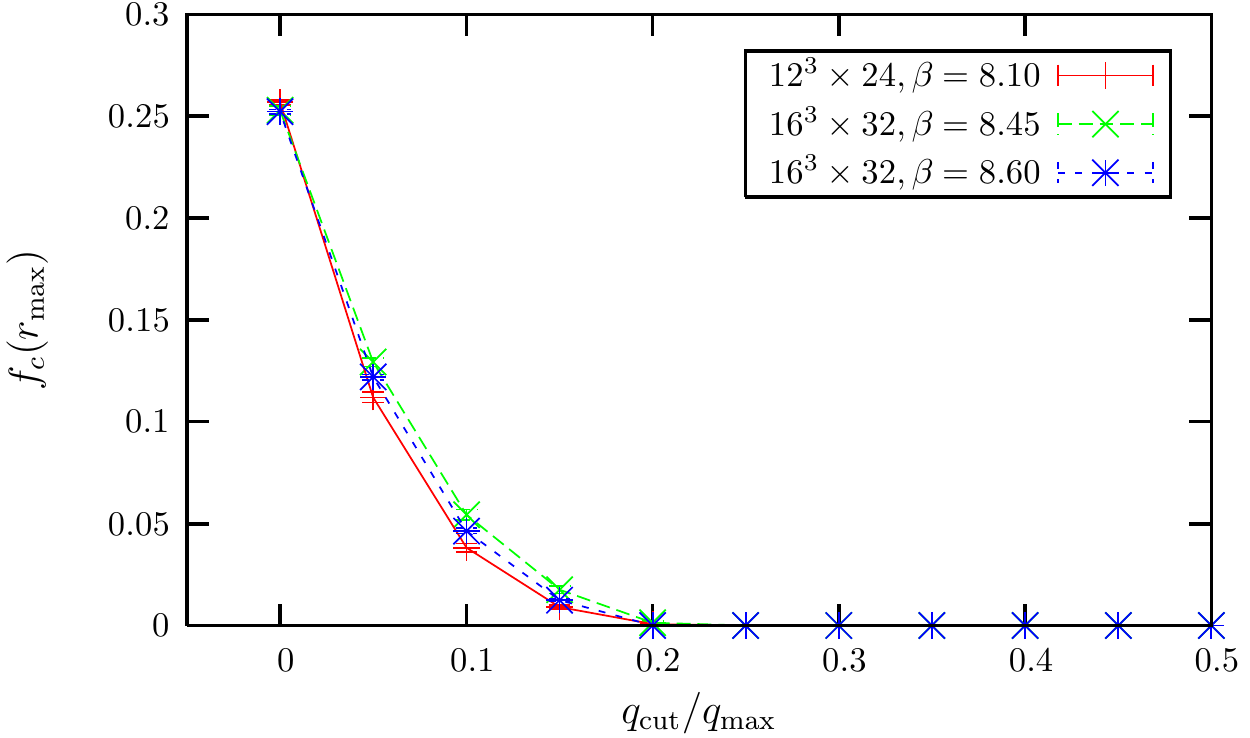}
\label{fig:refconnectivity}
}
\caption{ Left side from multi-probing method, right side from all-scale $q(x)$ in Ref.~\cite{qx:structure3}. 1st row: the distance between the two largest clusters. 2nd row: the connectivity $f(r_{max})$ of the largest cluster. }
\label{fig:qxstructure2}
\end{figure}

\begin{multicols}{2}
In Fig.~\ref{fig:qxstructure2}, the distance $ d(c, c^{\prime}) \approx \sqrt{3}a $ between two largest clusters at $q_{cut}=0$ is independent of $\beta$, like results shown in~\ref{fig:refdistance}, and just as the positive core of two-point correlator of topological charge density does.  This means that at $q_{cut}=0$  the two largest clusters of opposite sign  tangle with each other closely, when $a \rightarrow 0 $, they are just like two tangled  $3D$ papers dominating the whole $4D$ space. After the onset of percolation, the distance between two largest clusters reaches a plateau, which is also independent of $\beta$, this had been proved in Fig.~\ref{fig:refdistance} from Ref.~\cite{qx:structure3}. The turning point of cluster multiplicity, fractional size of the largest cluster and the plateau of the distance between two largest clusters all happen near the onset of percolation at $q_{cut} \approx 0.2 q_{max}$.

\end{multicols}
\begin{figure}[!htb]
\centering
\subfigure[$16^{3}\times32$, $a$=0.1161fm]{
\includegraphics[width=0.45\textwidth]{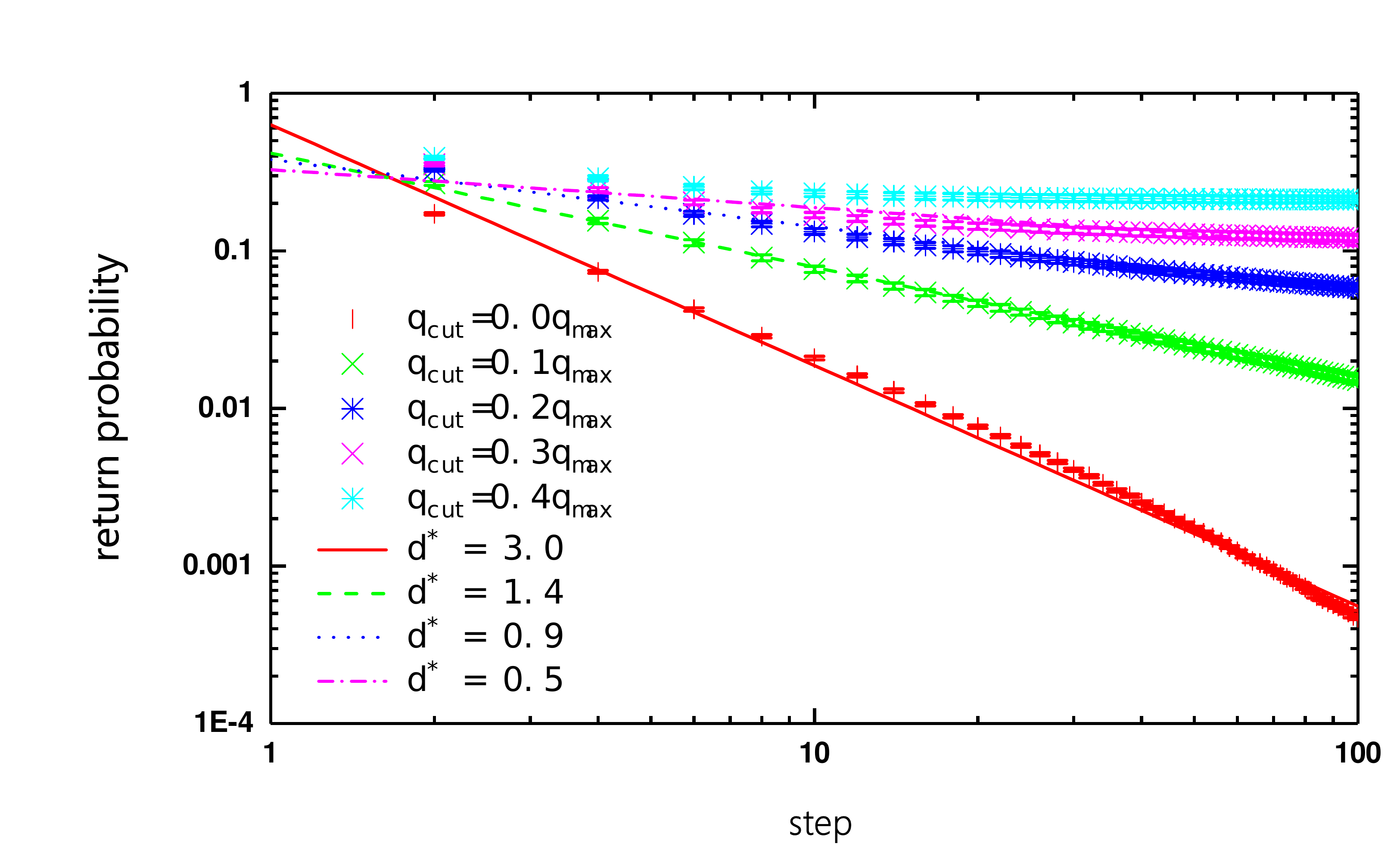}
\label{fig:dim9.0}
}
\subfigure[$16^{3}\times32$, $a$=0.0963fm]{
\includegraphics[width=0.45\textwidth]{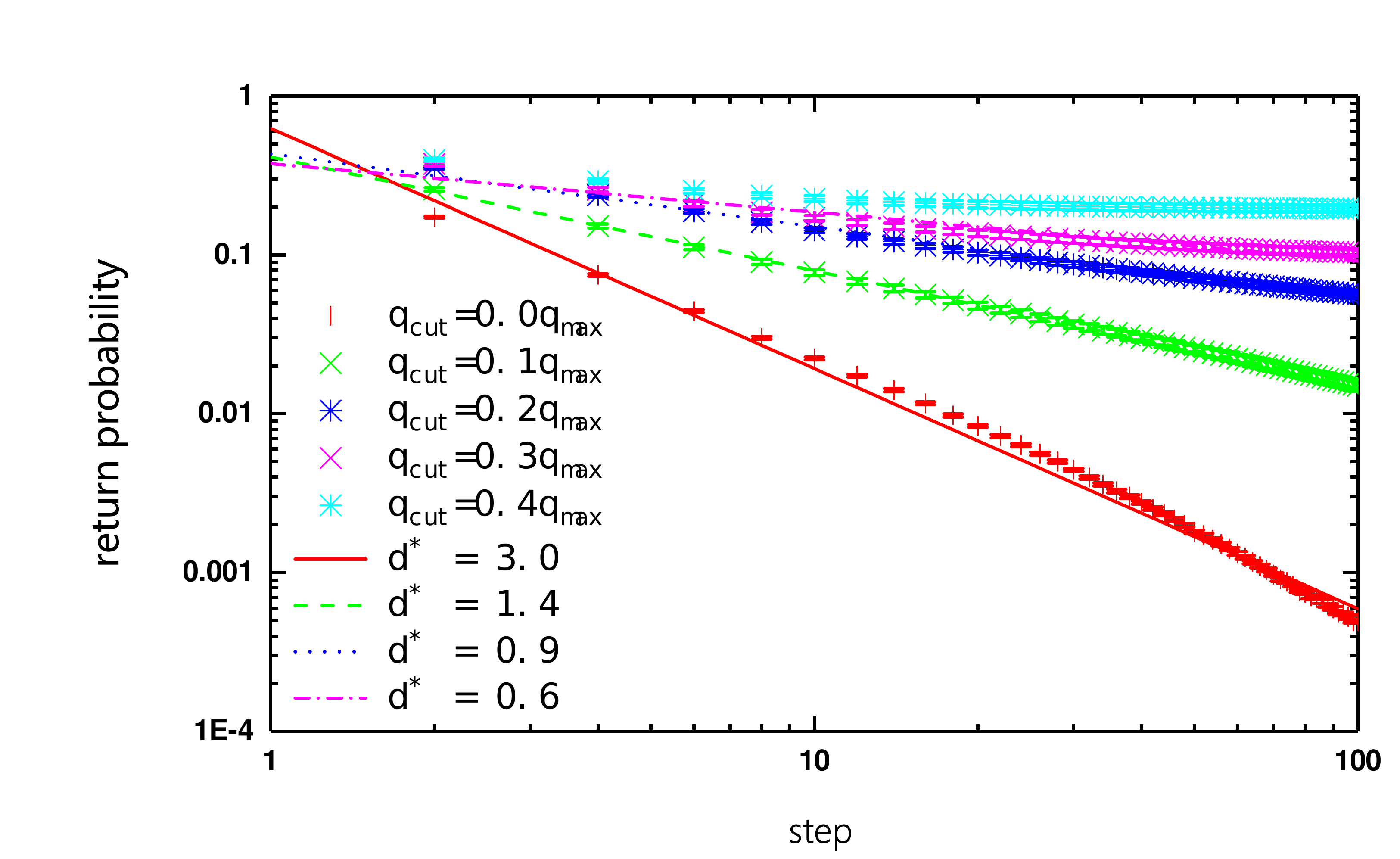}
\label{fig:dim9.4}
}
\subfigure[$16^{3}\times32$, $a$=0.0769fm]{
\includegraphics*[width=0.45\textwidth]{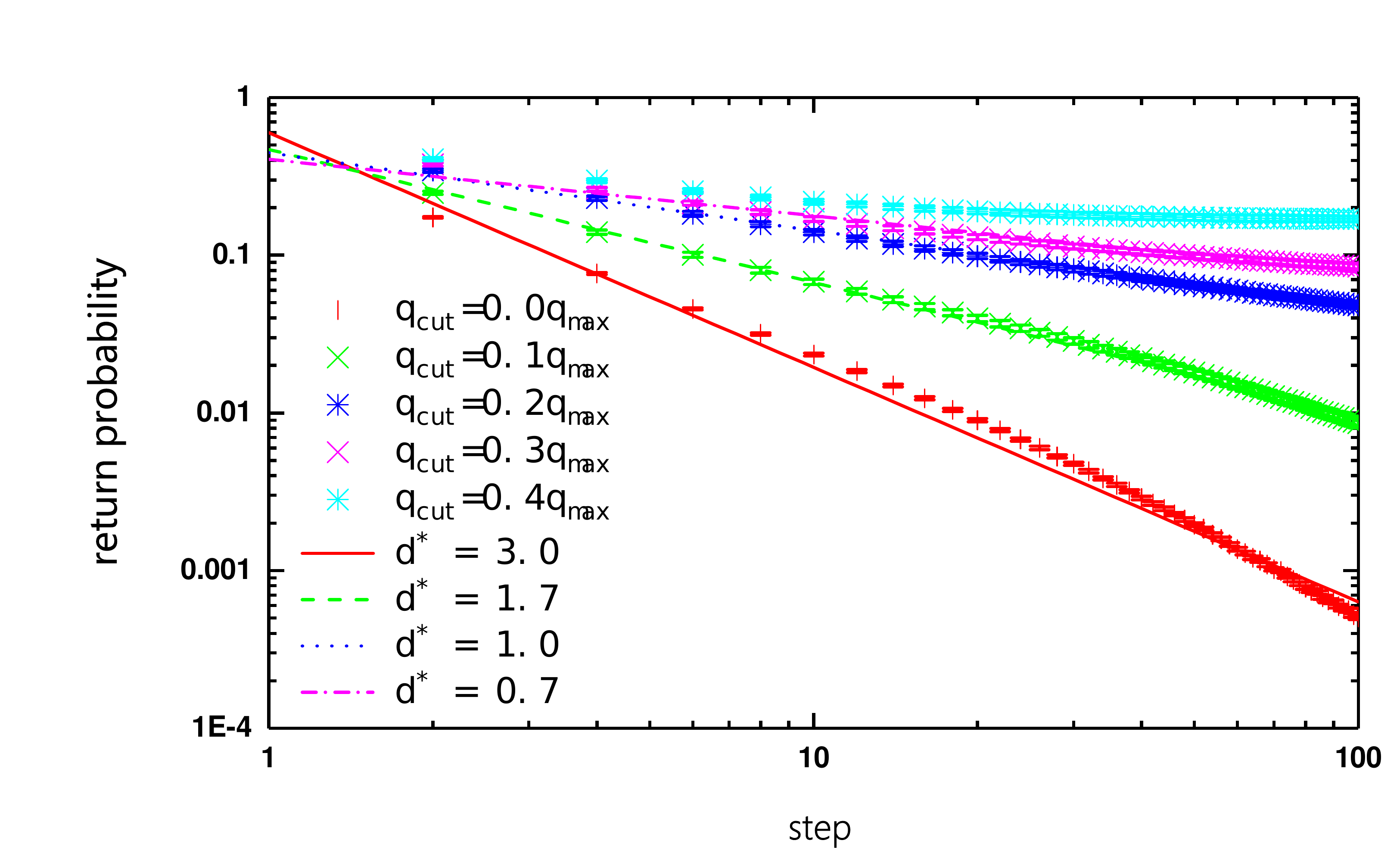}
\label{fig:dim10.0}
}
\subfigure[$12^{3}\times24$, $\beta=8.10$, from Ref.~\cite{qx:structure3}]{
\includegraphics*[viewport=4cm 15.8cm 17.5cm 23.5cm, width=0.45\textwidth]{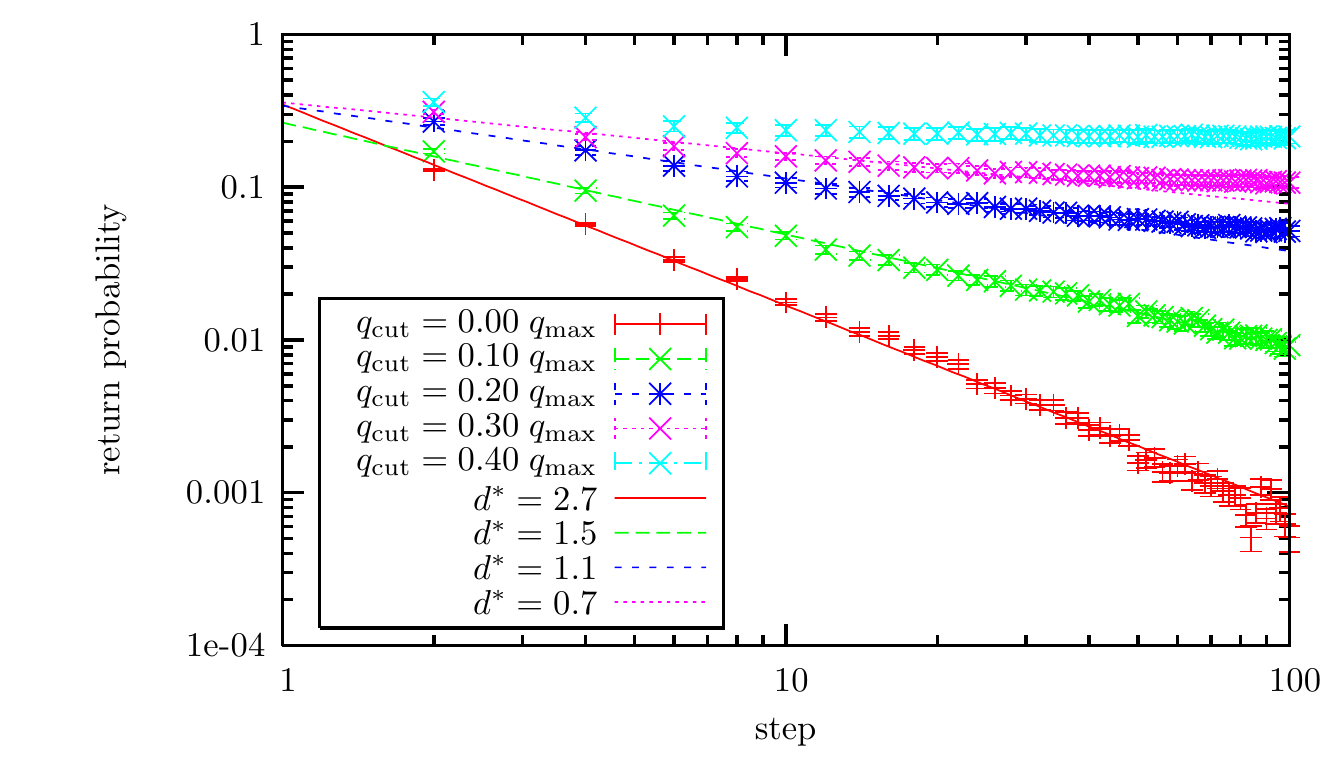}
\label{fig:refdim}
}
\caption{ Effective fractal dimension of the largest cluster, lines of $d^{\ast}$  are results of the power fits.}
\label{fig:qxdim}
\end{figure}

\begin{multicols}{2}
In Fig.~\ref{fig:qxdim} we present effective fractal dimension of the  largest cluster using random walker algorithm, lower than $4$ dimensional structure of topological charge density can be found. As the $q_{cut}$ increases, the effective fractal dimension drops down, but even  $q_{cut}$ at $0.20q_{max} - 0.25 q_{max}$, when  percolation is just starting, global clusters with lower effective dimension  $d^{\ast} = 1 - 1.4 $ have already appeared,  which is consistent with result from all-scale $q(x)$ showed in Fig.~\ref{fig:refdim}, and this structure had been first found as one dimensional skeleton in Ref.~\cite{qx:structure2}.

The trends of the properties of clusters with respect to $q_{cut}/q_{max}$ from our work are shown on left side of Figs.~\ref{fig:qxstructure1},~\ref{fig:qxstructure2},~\ref{fig:qxdim} are consistent with the trends shown on
the right sides in the same corresponding figures which we quote from Ref.~\cite{qx:structure3}. Most of the
 quantitative values are also compatible, except for the cluster multiplicity. Especially when $q_{cut} = 0$  there are still dozens of clusters survive, meanwhile in all-scale $q(x)$ there will be only two global clusters with opposite sign, and noting that the magnitude of cluster multiplicity in Fig.~\ref{fig:clusters} is about twice in Fig.~\ref{fig:refclusters}, which shows a scene much more chaotic than all-scale $q(x)$ does, this extra chaos or clusters might come from the multi-probing approximation method itself. After doing more study on the data and considering Fig.~\ref{fig:frac} is consistent with Fig.~\ref{fig:reffrac}, we found that these extra small clusters add to the total number of clusters, but they do not affect the fraction of the largest cluster to the total occupied volume substantially. Especially at $q_{cut}=0$, there are few small fragments composed of only one or two sites,  the total volume of these fragments can be ignored with respect to the whole lattice space or the two largest clusters with opposite sign, which are the clusters that all-scale $q(x)$ only have at $q_{cut}=0$.


Noting that in Figs.~\ref{fig:refclusters},~\ref{fig:reffrac},~\ref{fig:refdistance},~\ref{fig:refconnectivity}, lattice ensembles at $\beta = 8.10$ with $53$ configurations, at $\beta = 8.45$ with $5$ configurations, and $2$ configurations at $\beta = 8.60$, the numbers of configurations are so small owing to that the direct all-scale $q(x)$ calculating consumes too much computer resource. All the $3$ ensembles used in this work consist of $200$ configurations.

\end{multicols}
\begin{figure}[!htb]
\subfigure[Ensemble $16^{3}\times32$, $a$=0.1161fm]{
\includegraphics[width=0.30\textwidth, height=0.20\textwidth]{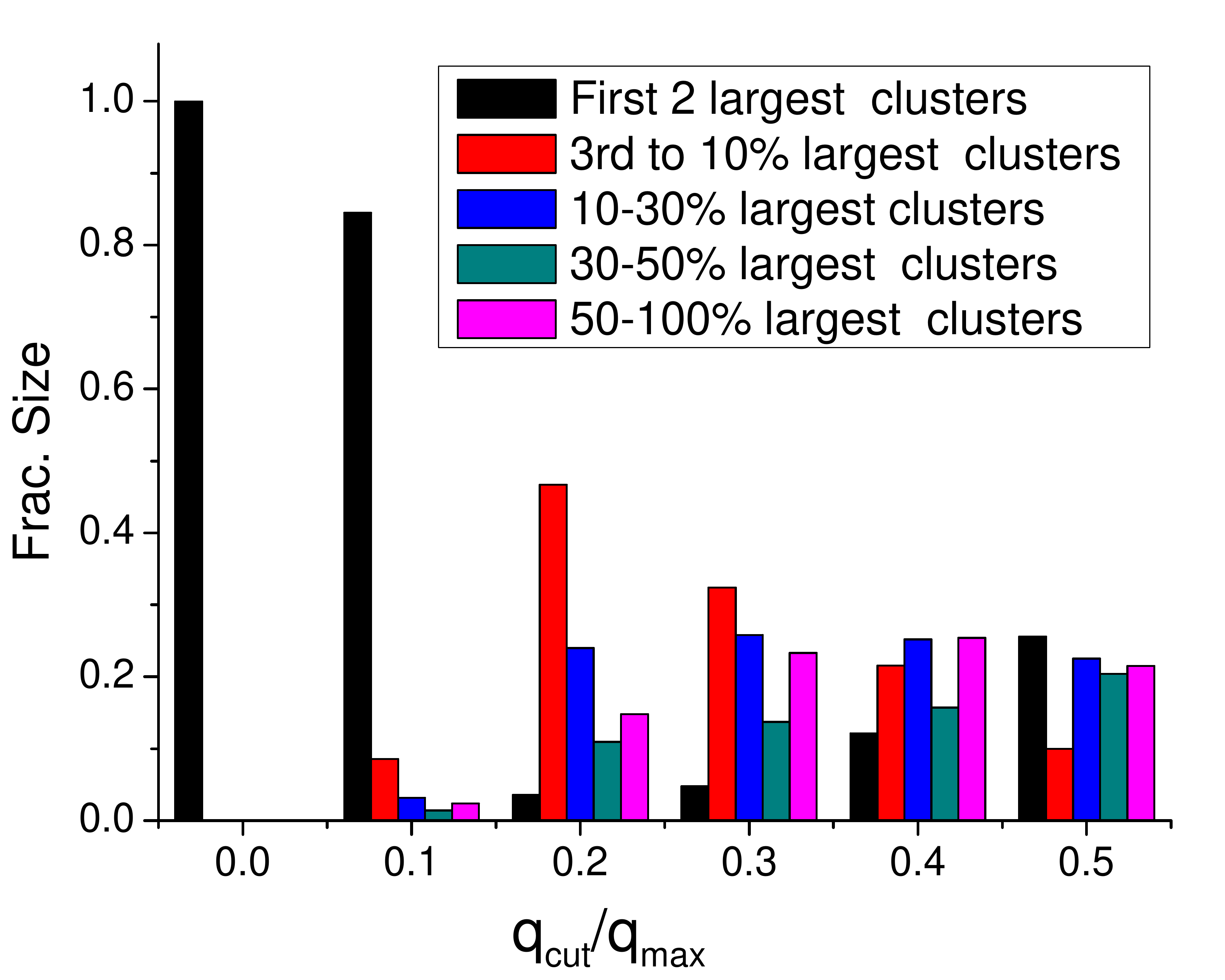}
\label{fig:distr9.0}
}
\subfigure[Ensemble $16^{3}\times32$, $a$=0.0963fm]{
\includegraphics[width=0.30\textwidth, height=0.20\textwidth]{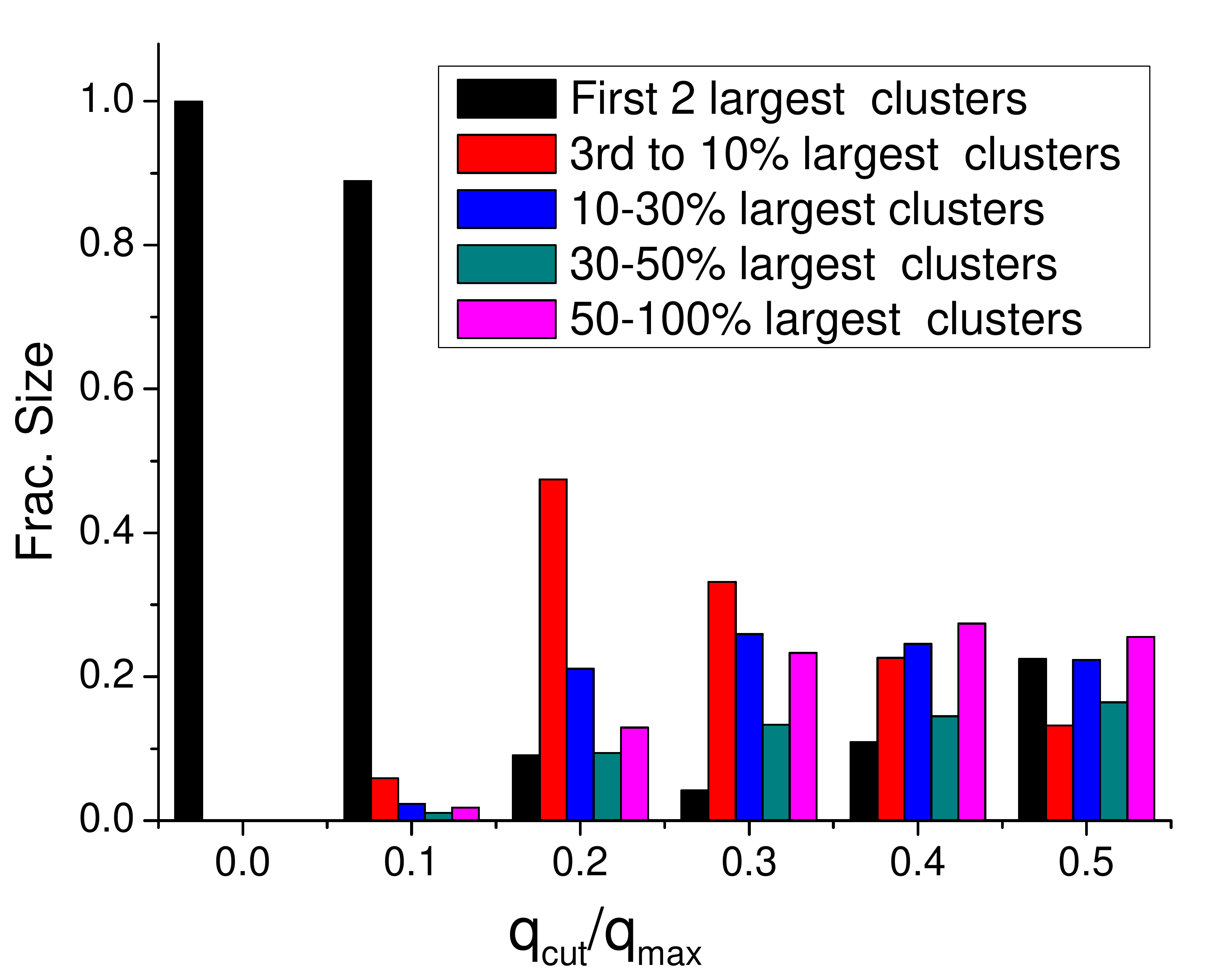}
\label{fig:distr9.4}
}
\subfigure[Ensemble $16^{3}\times32$, $a$=0.0769fm]{
\includegraphics[width=0.30\textwidth, height=0.20\textwidth]{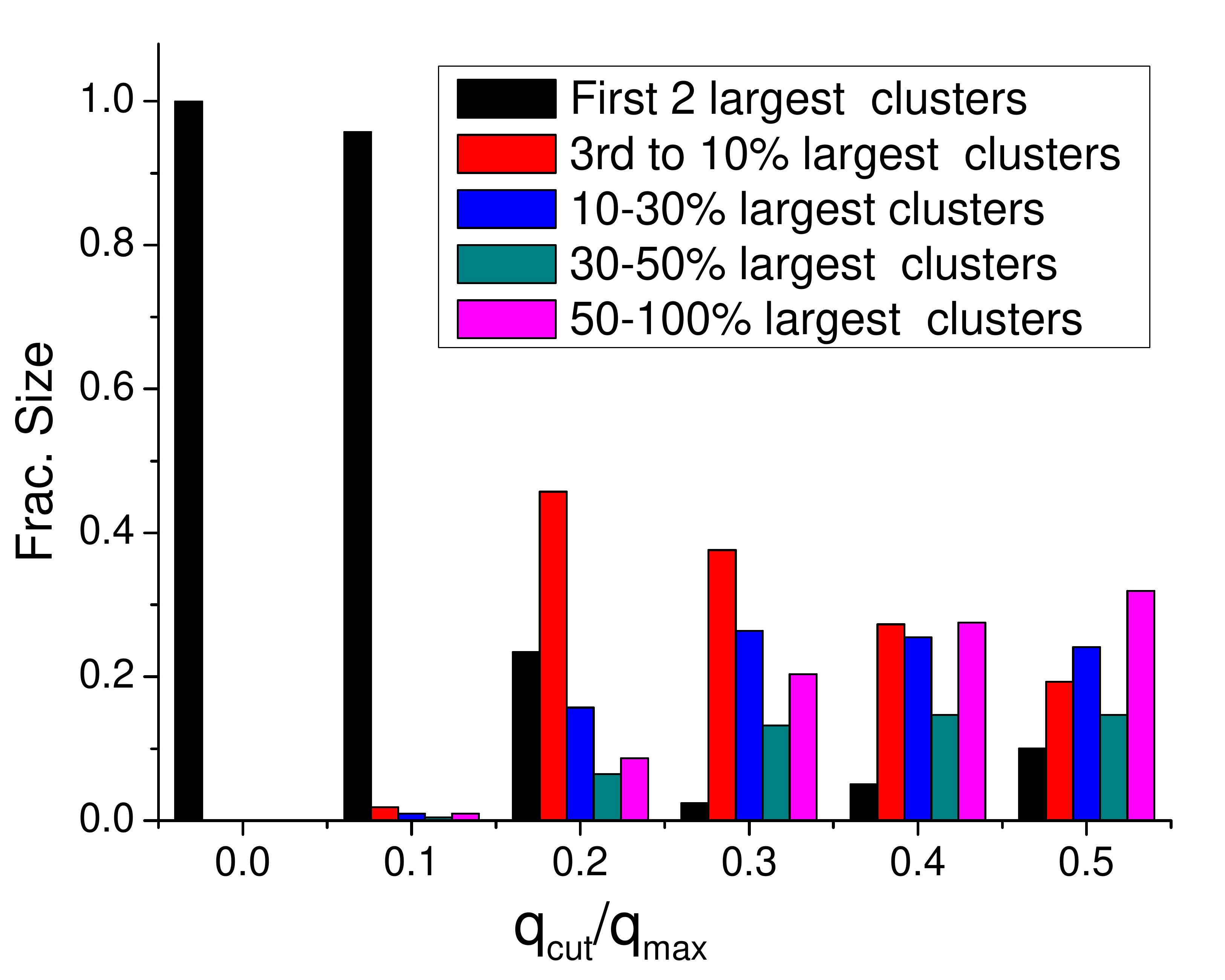}
\label{fig:distr10.0}
}
\caption{ The average fractional size of largest clusters  relative to occupied volume by total clusters, separated by sections of size of clusters: the first 2 largest clusters, the 3rd to 10\% largest clusters, 10\% to 30\% largest clusters, 30\% to 50\% largest clusters, and the left 50\%.}
\label{fig:distribution_frac}
\end{figure}

\begin{multicols}{2}
In Fig.~\ref{fig:distribution_frac} we present the fractional size, which is separated by different sections of size of clusters, relative to the total occupied  volume by clusters, and the clusters are ranked according to the volume of clusters. From Fig.~\ref{fig:distribution_frac} we can find that most volume are occupied by the first half largest clusters in three different ensembles below the onset of the percolation $q_{cut} = 0.2q_{max}$, then the fractional size of extra clusters should not be concerned. But according to Figs.~\ref{fig:qxstructure1},~\ref{fig:qxstructure2},~\ref{fig:qxdim}, above the onset of percolation the extra clusters from multi-probing approximation still do not influence the role of the largest clusters substantially.

It is easy to understand the reason why multi-probing method will bring extra clusters. Because we use the clusters of topological charge density as tools to investigate the structures of the quenched QCD vacuum, with the hope that the structures would reflect ordered properties of $q(x)$, and the multi-probing method will bring superposition from different probing sites to topological charge density, which can be regarded as disordered disturbance, then noise will be introduced to the ordered structure of clusters.

Noting that the properties of clusters of topological charge density from multi-probing method showed above are much closer to the all-scale topological charge density than that from eigenmode expansion showed in Figs.~20,~21 from Ref.~\cite{qx:structure3}.

\subsection{ Random permutation of topological charge density }

Though we have multi-probing observed structures of topological charge density,
it is still a question where these stuctures come from.
For example, are these structures arise from the ordered properties
of the quenched QCD vacuum or they can be established by disorders.
To get some information about this, we permute the topological charge density
randomly through space-time to generate stochastically distributed $q(x)$, which should no longer contain information of the ordered properties
of the quenched QCD vacuum, and then investigate the consequences of
this permutation.

\end{multicols}
\begin{figure}[!hbt]
\centering
\subfigure[]{
\includegraphics[width=0.40\textwidth, height=0.25\textwidth]{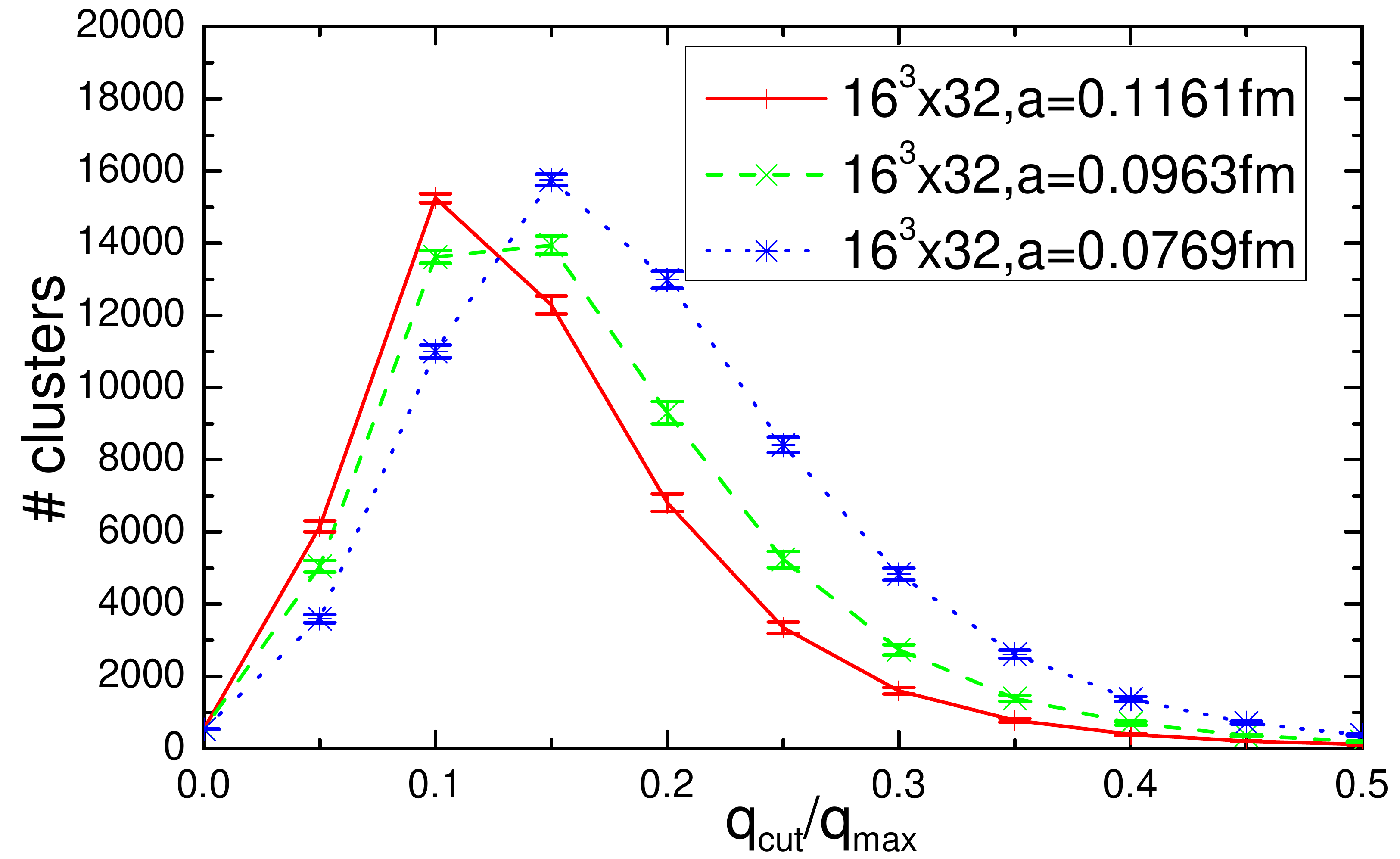}
\label{fig:rp_clusters}
}
\subfigure[]{
\includegraphics[width=0.40\textwidth, height=0.25\textwidth]{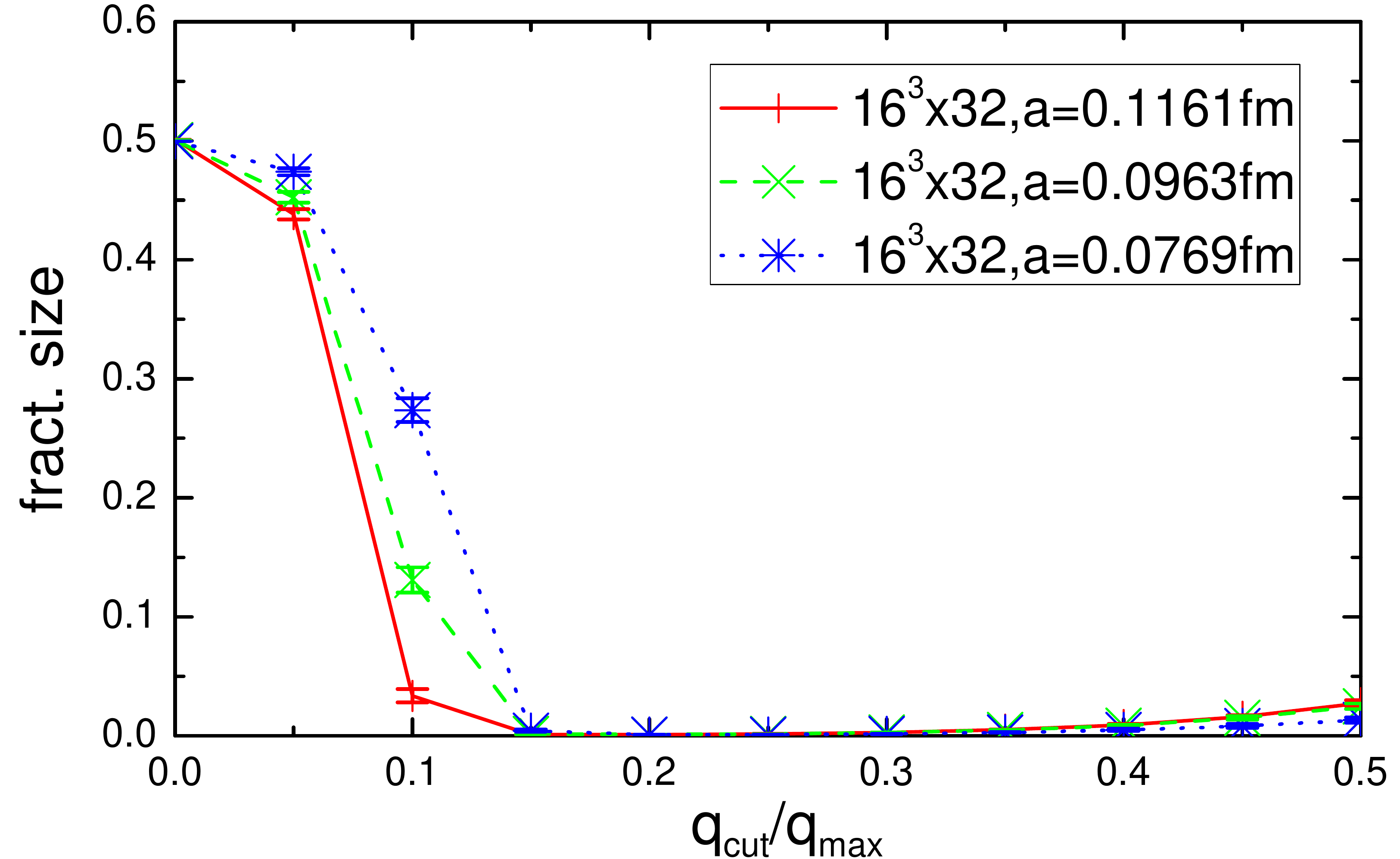}
\label{fig:rp_frac}
}
\subfigure[]{
\includegraphics[width=0.40\textwidth, height=0.25\textwidth]{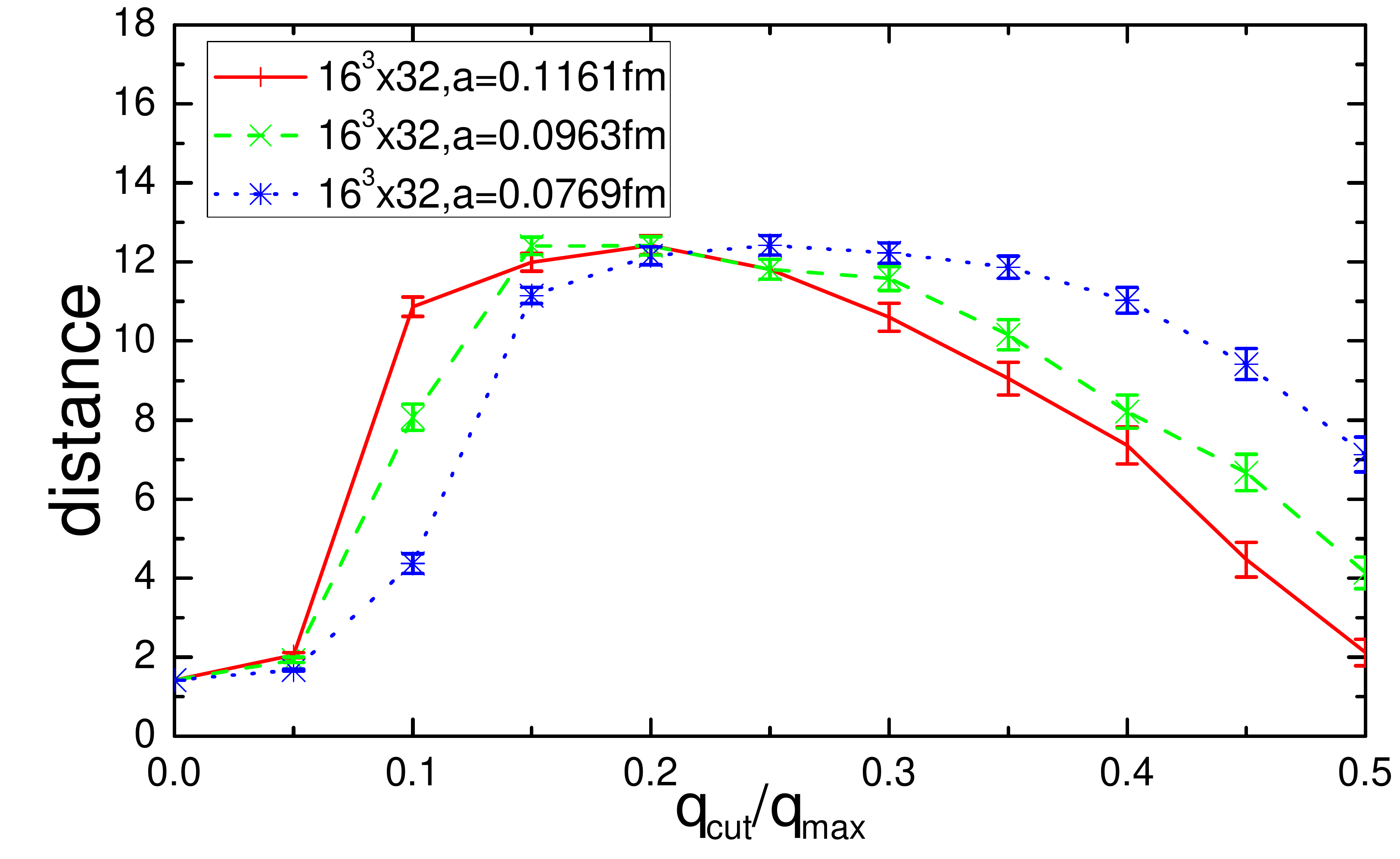}
\label{fig:rp_distance}
}
\subfigure[]{
\includegraphics[width=0.40\textwidth, height=0.25\textwidth]{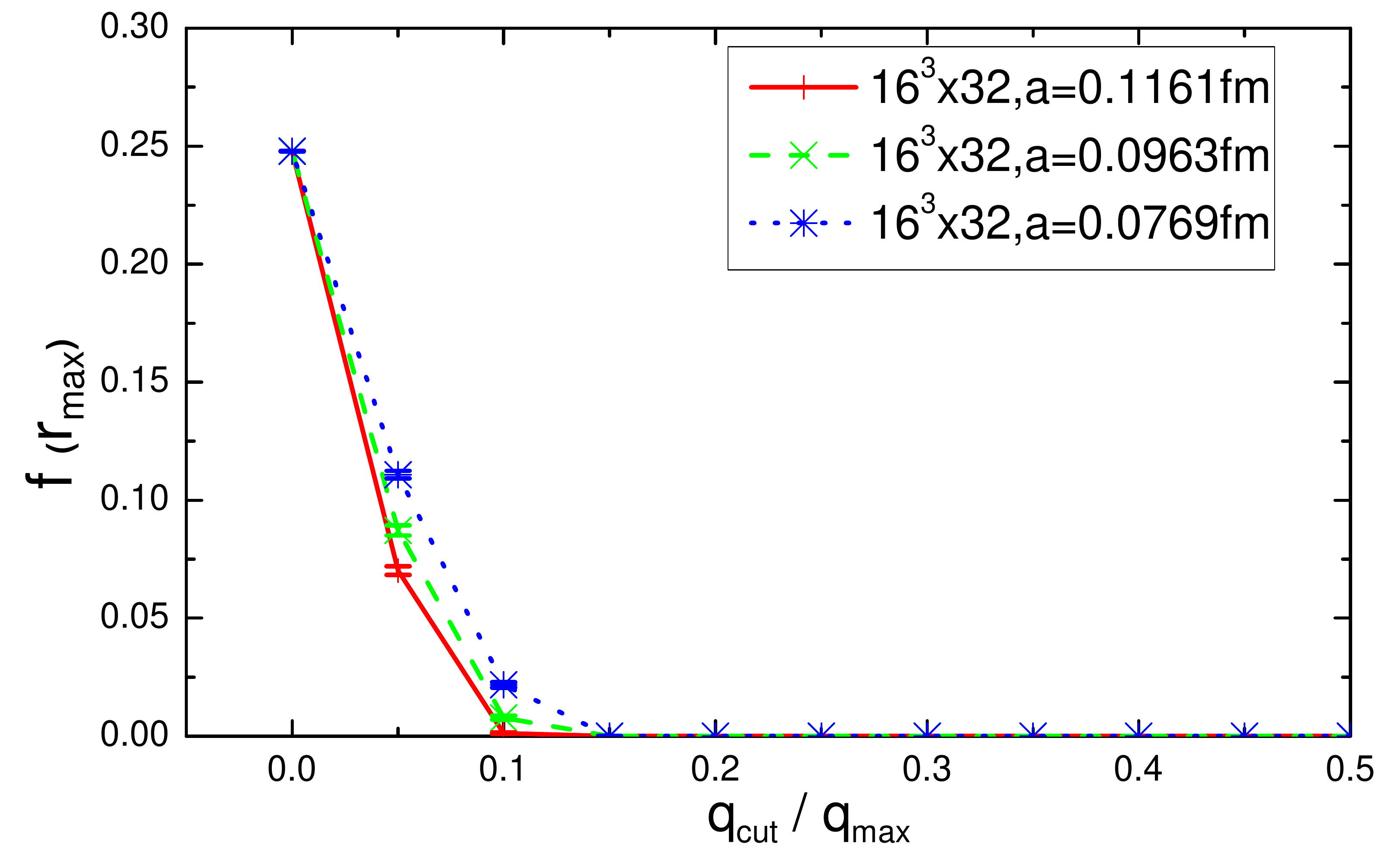}
\label{fig:rp_connectivity}
}
\subfigure[$16^{3}\times32$, $a$=0.1161fm]{
\includegraphics[width=0.31\textwidth, height=0.20\textwidth]{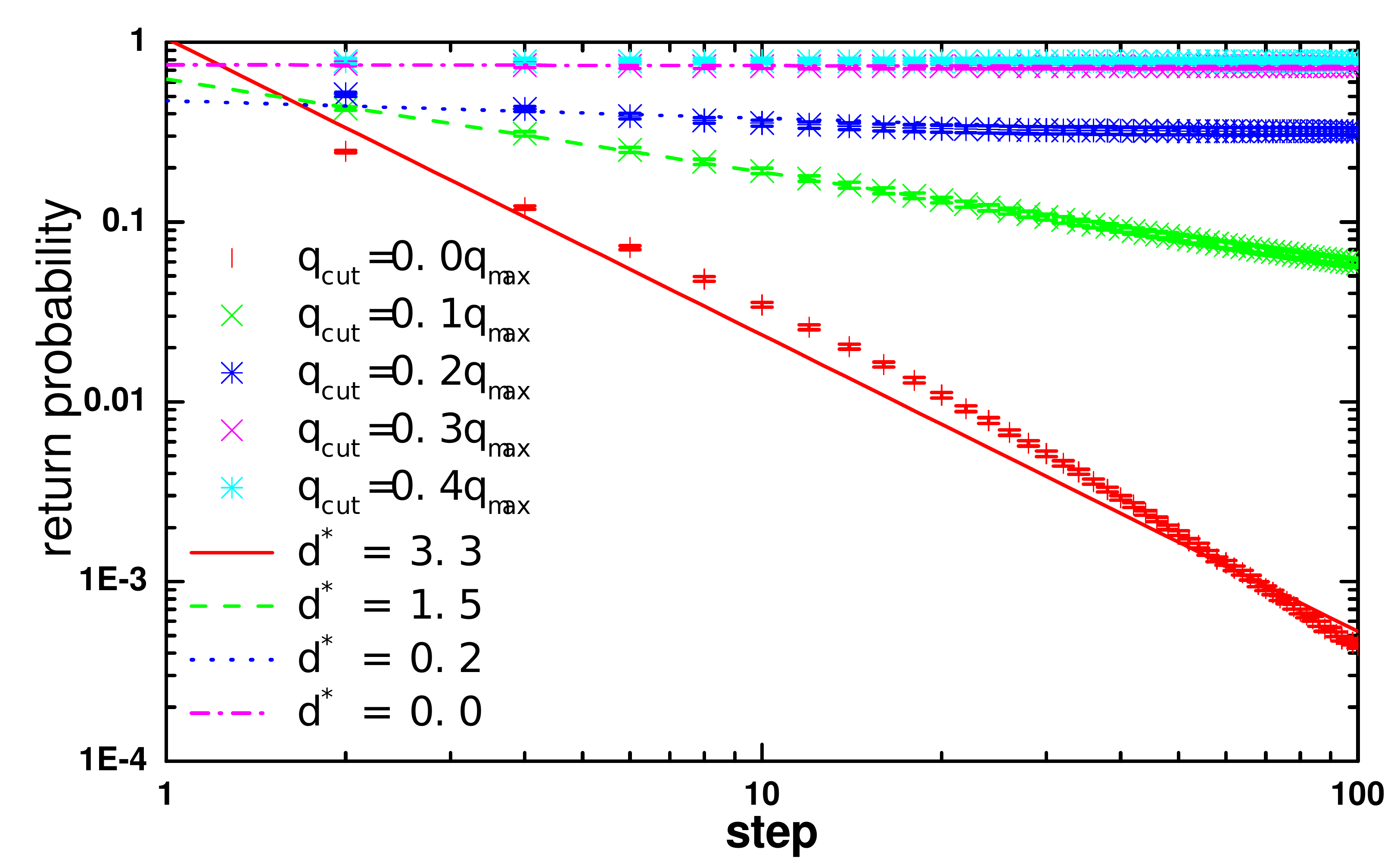}
\label{fig:rp_dim9.0}
}
\subfigure[$16^{3}\times32$, $a$=0.0963fm]{
\includegraphics[width=0.31\textwidth, height=0.20\textwidth]{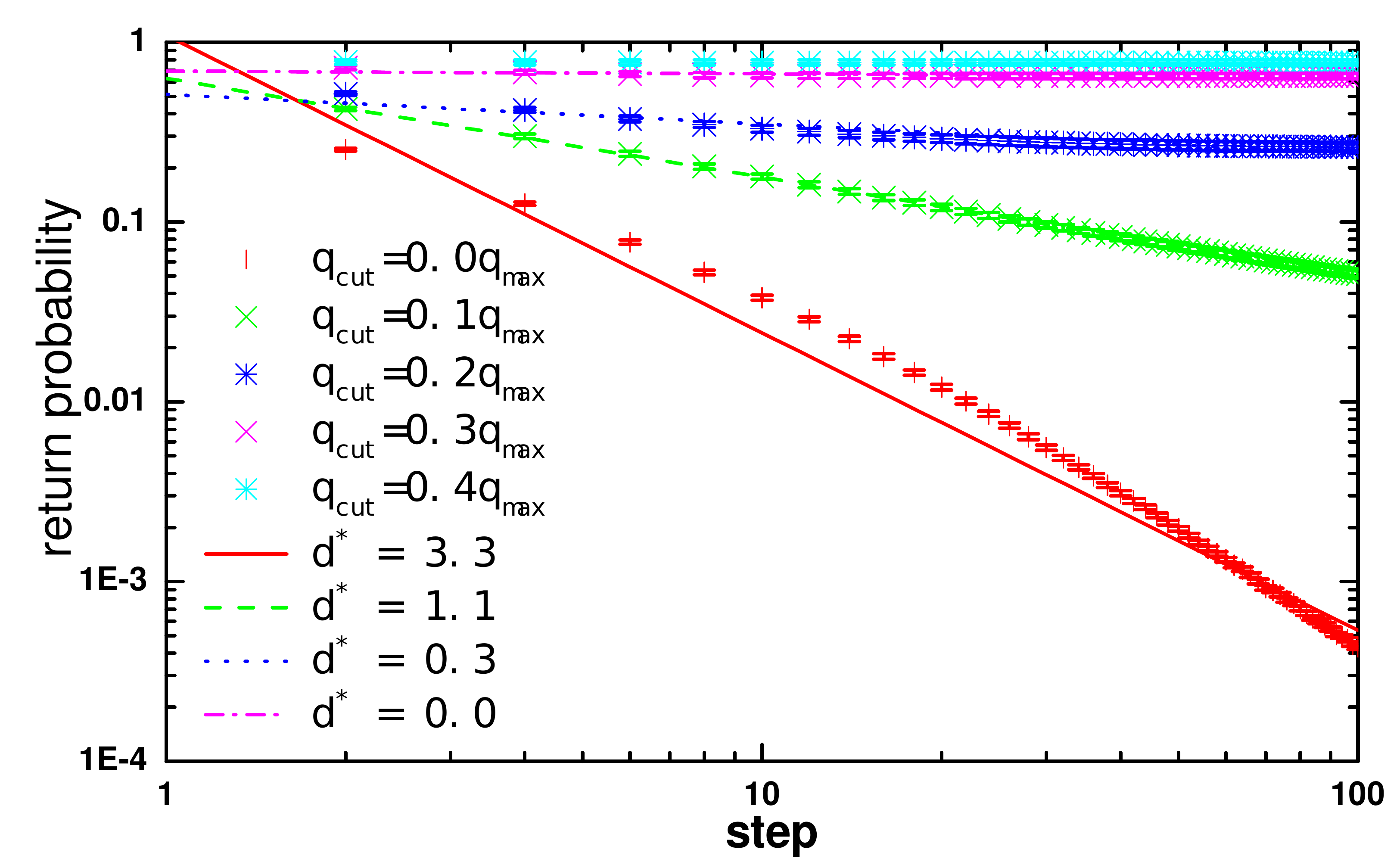}
\label{fig:rp_dim9.4}
}
\subfigure[$16^{3}\times32$, $a$=0.0769fm]{
\includegraphics[width=0.31\textwidth, height=0.20\textwidth]{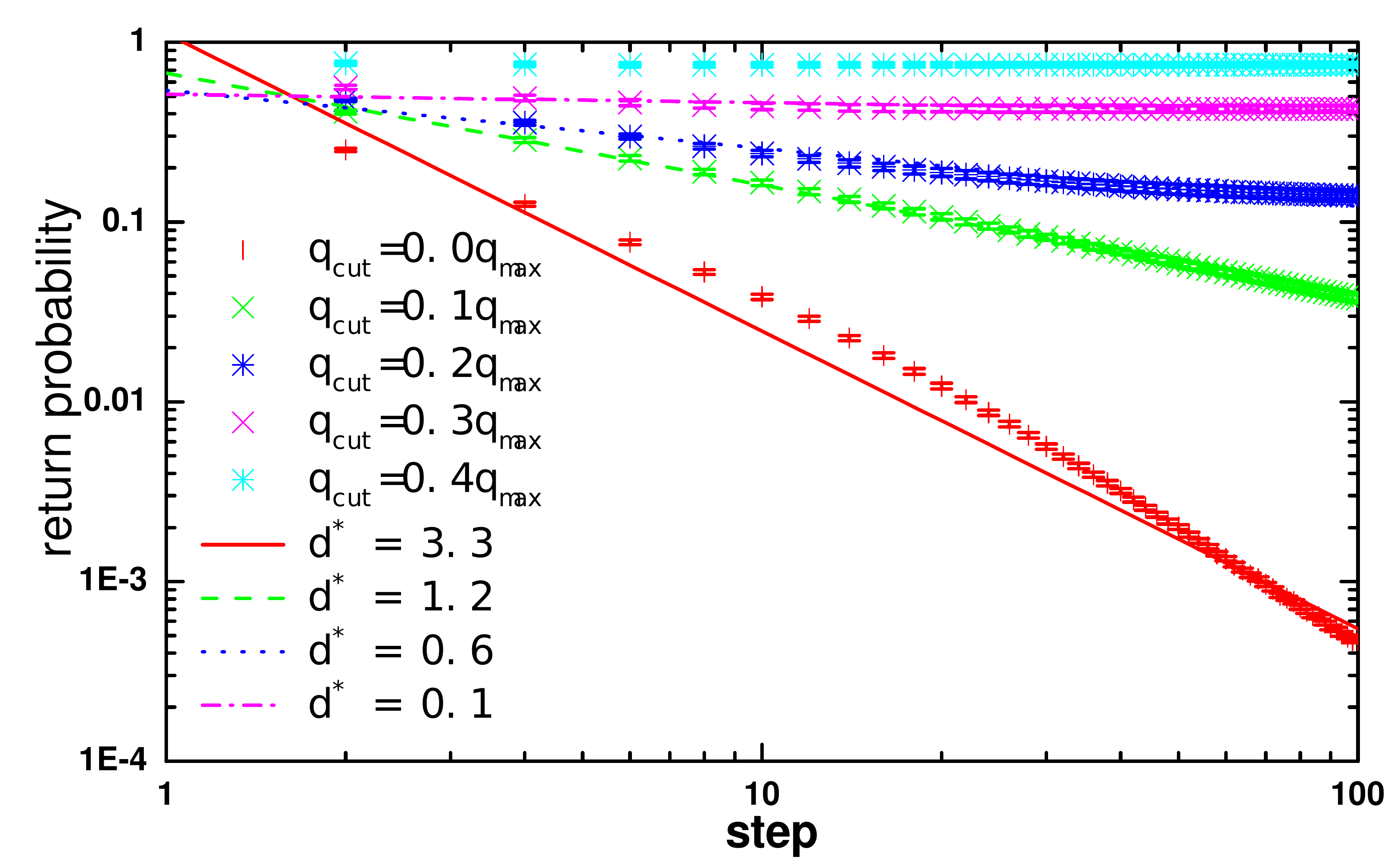}
\label{fig:rp_dim10.0}
}
\caption{ Properties of clusters structure of permuted topological charge density.
\ref{fig:rp_clusters}: the cluster multiplicity,~\ref{fig:rp_frac}: the  size of the largest cluster relative to all clusters,~\ref{fig:rp_distance}: the distance between the two largest clusters,~\ref{fig:rp_connectivity}: the connectivity $f(r_{max})$ of the largest cluster,~\ref{fig:rp_dim9.0}~\ref{fig:rp_dim9.4}~\ref{fig:rp_dim10.0}: the returning probability of the largest cluster, lines of $d^{\ast}$  are results of the power fits.}
\label{fig:rp}
\end{figure}

\begin{multicols}{2}
Fig.~\ref{fig:rp} shows results of the permuted topological charge density.
It is found that the cluster multiplicity becomes much larger than in Fig.~\ref{fig:clusters}.
Even at $q_{cut}=0$ there are still hundreds of clusters.
The onset of percolation is  found near $0.15q_{max}$ now, the fractional size of the largest cluster and the connectivity drop quickly, the distance are $\sqrt{2}a$, which is less than results above in Fig.~\ref{fig:distance}. These results suggest that the clusters tangle with each other more closely. Above the onset of percolation, the distance between the two largest clusters drops down, when it should reach a plateau before. The effective dimension shown in Figs.~\ref{fig:rp_dim9.0},~\ref{fig:rp_dim9.4},~\ref{fig:rp_dim10.0} drops faster than those shown in Figs.~\ref{fig:dim9.0},~\ref{fig:dim9.4},~\ref{fig:dim10.0} and become negligible  at $q_{cut}=0.4q_{max}$ already. On the other hand, the changes shown in Fig.~\ref{fig:rp} also demonstrate that, the differences between results of multi-probing and all-scale $q(x)$ seen above really come from disordered disturbance on the lattice. Therefore we can be sure to say that the properties  of clusters from multi-probing method do reflect the underlying ordered structure of QCD vacuum, especially the structure recognized between $q_{cut}=0.15q_{max} - 0.20q_{max} $ where the structure of clusters is similar to all-scale $q(x)$ but different from the permuted case.

\section{Summary}

 In this paper, using the multi-probing method with overlap fermion to approach the topological charge density,
 the topological aspects of quenched QCD are investigated. To be specific, we study the
 topological charge, topological susceptibility and the structure of topological charge density in space-time.
 It is found that the cluster properties are consistent with the results obtained from all-scale topological charge density rather well. After comparing them with clusters of randomly permuted topological charge, we conclude that, the global structure that appear near the onset of percolation with effective dimension near $1$ may well reflect the underlying ordered properties of the quenched QCD vacuum. Since there are still global clusters survive when the sites that have topological charge density of absolute value below $q_{cut}$ have been discarded.

\section*{Acknowledgments}

This work was mainly run on Tianhe-2 supercomputer at NSCC in Guangzhou.

\section{References}

\bibliographystyle{unsrt}
\bibliography{ref}

\end{multicols}

\clearpage
\end{document}